\documentclass{entcs}
\usepackage{entcsmacro}
\usepackage[matrix,frame,arrow]{xy}

\makeatletter

\usepackage{amstext}
\usepackage{amssymb}
\usepackage{stmaryrd}
\DeclareFontFamily{OT1}{cmtex}{}
\DeclareFontShape{OT1}{cmtex}{m}{n}
  {<5><6><7><8>cmtex8
   <9>cmtex9
   <10><10.95><12><14.4><17.28><20.74><24.88>cmtex10}{}
\DeclareFontShape{OT1}{cmtex}{m}{it}
  {<-> ssub * cmtt/m/it}{}

\DeclareFontShape{OT1}{cmtt}{bx}{n}
  {<5><6><7><8>cmtt8
   <9>cmbtt9
   <10><10.95><12><14.4><17.28><20.74><24.88>cmbtt10}{}
\DeclareFontShape{OT1}{cmtex}{bx}{n}
  {<-> ssub * cmtt/bx/n}{}

\newcommand{\Conid}[1]{\mathit{#1}}
\newcommand{\Varid}[1]{\mathit{#1}}
\newcommand{\anonymous}{\kern0.06em \vbox{\hrule\@width.5em}}
\newcommand{\plus}{\mathbin{+\!\!\!+}}
\newcommand{\bind}{\mathbin{>\!\!\!>\mkern-6.7mu=}}

\usepackage{polytable}

\@ifundefined{mathindent}%
  {\newdimen\mathindent\mathindent\leftmargini}%
  {}%

\def\resethooks{%
  \global\let\SaveRestoreHook\empty
  \global\let\ColumnHook\empty}
\newcommand*{\savecolumns}[1][default]%
  {\g@addto@macro\SaveRestoreHook{\savecolumns[#1]}}
\newcommand*{\restorecolumns}[1][default]%
  {\g@addto@macro\SaveRestoreHook{\restorecolumns[#1]}}
\newcommand*{\aligncolumn}[2]%
  {\g@addto@macro\ColumnHook{\column{#1}{#2}}}

\resethooks

\newcommand{\onelinecommentchars}{\quad-{}- }
\newcommand{\commentbeginchars}{\enskip\{-}
\newcommand{\commentendchars}{-\}\enskip}

\newcommand{\visiblecomments}{%
  \let\onelinecomment=\onelinecommentchars
  \let\commentbegin=\commentbeginchars
  \let\commentend=\commentendchars}

\newcommand{\invisiblecomments}{%
  \let\onelinecomment=\empty
  \let\commentbegin=\empty
  \let\commentend=\empty}

\visiblecomments

\newlength{\blanklineskip}
\newcommand{\hsindent}[1]{\quad}

\makeatother

\newcommand{\Qo}{\mathbf{Q}^\circ}
\newcommand{\QVec}{\mathbf{Vec}}

\newcommand{\Qom}[2]{\Qo\,#1\,#2}

\newcommand{\Vec}[1]{\mathbf{V}~#1}
\newcommand{\ip}[2]{\langle #1 | #2 \rangle}
\newcommand{\vreturn}[1]{\textit{return}~#1}
\newcommand{\vconst}[1]{\textit{const}~#1}

\newcommand{\Complex}{\mathbb{C}}

\newcommand{\RealPlus}{\mathbb{R}^+}
\renewcommand{\conj}[1]{{#1}^*}

\newcommand{\ot}{\otimes}
\newcommand{\emptyC}{\bullet}
\newcommand{\G}{\Gamma}
\newcommand{\D}{\Delta}
\newcommand{\dom}[1]{\textrm{dom}\,{#1}}

\newcommand{\true}{\mathrm{true}}
\newcommand{\false}{\mathrm{false}}

\newcommand{\vdasho}{\vdash^\circ}

\newcommand{\ru}[2]{\vspace{1ex}
\begin{prooftree}
#1 \justifies #2
\end{prooftree}\vspace{1ex}}

\newcommand{\Ru}[3]{\vspace{1ex}
\begin{prooftree}
#1 \justifies #2 \using{\rm{#3}}
\end{prooftree}\vspace{1ex}}
\newcommand{\Ax}[2]{
\Ru{}{#1}{#2} }

\newcommand{\rin}{\ \mathtt{in}\ }

\newcommand{\rifo}{\ \mathtt{if}^{\circ}\ \ }

\newcommand{\rthen}{\ \ \mathtt{then}\ \ }
\newcommand{\relse}{\ \ \mathtt{else}\ \ }
\newcommand{\rlet}{\mathtt{let}\ }

\newcommand{\tin}{:}

\newcommand{\cond}[2]{#1 | #2}
\newcommand{\condo}[2]{#1 |^\circ #2}

\newcommand{\evalC}[1]{\llbracket{#1}\rrbracket}
\newcommand{\evalV}[1]{\llbracket{#1}\rrbracket^\textrm{Q}}
\newcommand{\evalQ}[1]{\llbracket{#1}\rrbracket^\textrm{Q}}
\newcommand{\evalP}[1]{\llbracket{#1}\rrbracket^\textrm{P}}

\newcommand{\evalx}[1]{\llbracket{#1}\rrbracket}

\newcommand{\ValC}{\mathrm{Val}^\mathrm{C}}
\newcommand{\ValQ}{\mathrm{Val}^\mathrm{Q}}
\newcommand{\ValQo}{\mathrm{Val}_\circ^\mathrm{Q}}
\newcommand{\val}{\mathrm{val}}
\newcommand{\fst}{\mathrm{fst}}

\newcommand{\pinv}[1]{1/{#1}}
\newcommand{\con}[1]{|#1|}
\newcommand{\Tm}{\mathrm{Tm}}
\newcommand{\CTm}[1]{\Tm\,#1}
\newcommand{\Tmm}[2]{\Tm\,#1\,#2}
\newcommand{\deltax}{\hat{\delta}}

\newcommand{\vecx}[1]{{\overrightarrow{#1}}}

\newdimen\proofrulebreadth \proofrulebreadth=.05em
\newdimen\proofdotseparation \proofdotseparation=1.25ex
\newdimen\proofrulebaseline \proofrulebaseline=2ex
\newcount\proofdotnumber \proofdotnumber=3
\let\then\relax
\def\hfi{\hskip0pt plus.0001fil}
\mathchardef\squigto="3A3B
\newif\ifinsideprooftree\insideprooftreefalse
\newif\ifonleftofproofrule\onleftofproofrulefalse
\newif\ifproofdots\proofdotsfalse
\newif\ifdoubleproof\doubleprooffalse
\let\wereinproofbit\relax
\newdimen\shortenproofleft
\newdimen\shortenproofright
\newdimen\proofbelowshift
\newbox\proofabove
\newbox\proofbelow
\newbox\proofrulename
\def\shiftproofbelow{\let\next\relax\afterassignment\setshiftproofbelow\dimen0 }
\def\shiftproofbelowneg{\def\next{\multiply\dimen0 by-1 }%
\afterassignment\setshiftproofbelow\dimen0 }
\def\setshiftproofbelow{\next\proofbelowshift=\dimen0 }
\def\setproofrulebreadth{\proofrulebreadth}

\def\prooftree{
\ifnum  \lastpenalty=1
\then   \unpenalty
\else   \onleftofproofrulefalse
\fi
\ifonleftofproofrule
\else   \ifinsideprooftree
        \then   \hskip.5em plus1fil
        \fi
\fi
\bgroup
\setbox\proofbelow=\hbox{}\setbox\proofrulename=\hbox{}%
\let\justifies\proofover\let\leadsto\proofoverdots\let\Justifies\proofoverdbl
\let\using\proofusing\let\[\prooftree
\ifinsideprooftree\let\]\endprooftree\fi
\proofdotsfalse\doubleprooffalse
\let\thickness\setproofrulebreadth
\let\shiftright\shiftproofbelow \let\shift\shiftproofbelow
\let\shiftleft\shiftproofbelowneg
\let\ifwasinsideprooftree\ifinsideprooftree
\insideprooftreetrue
\setbox\proofabove=\hbox\bgroup$\displaystyle 
\let\wereinproofbit\prooftree
\shortenproofleft=0pt \shortenproofright=0pt \proofbelowshift=0pt
\onleftofproofruletrue\penalty1
}

\def\eproofbit{
\ifx    \wereinproofbit\prooftree
\then   \ifcase \lastpenalty
        \then   \shortenproofright=0pt  
        \or     \unpenalty\hfil         
        \or     \unpenalty\unskip       
        \else   \shortenproofright=0pt  
        \fi
\fi
\global\dimen0=\shortenproofleft
\global\dimen1=\shortenproofright
\global\dimen2=\proofrulebreadth
\global\dimen3=\proofbelowshift
\global\dimen4=\proofdotseparation
\global\count255=\proofdotnumber
$\egroup  
\shortenproofleft=\dimen0
\shortenproofright=\dimen1
\proofrulebreadth=\dimen2
\proofbelowshift=\dimen3
\proofdotseparation=\dimen4
\proofdotnumber=\count255
}

\def\proofover{
\eproofbit 
\setbox\proofbelow=\hbox\bgroup 
\let\wereinproofbit\proofover
$\displaystyle
}%
\def\proofoverdbl{
\eproofbit 
\doubleprooftrue
\setbox\proofbelow=\hbox\bgroup 
\let\wereinproofbit\proofoverdbl
$\displaystyle
}%
\def\proofoverdots{
\eproofbit 
\proofdotstrue
\setbox\proofbelow=\hbox\bgroup 
\let\wereinproofbit\proofoverdots
$\displaystyle
}%
\def\proofusing{
\eproofbit 
\setbox\proofrulename=\hbox\bgroup 
\let\wereinproofbit\proofusing
\kern0.3em$
}

\def\endprooftree{
\eproofbit 
  \dimen5 =0pt
\dimen0=\wd\proofabove \advance\dimen0-\shortenproofleft
\advance\dimen0-\shortenproofright
\dimen1=.5\dimen0 \advance\dimen1-.5\wd\proofbelow
\dimen4=\dimen1
\advance\dimen1\proofbelowshift \advance\dimen4-\proofbelowshift
\ifdim  \dimen1<0pt
\then   \advance\shortenproofleft\dimen1
        \advance\dimen0-\dimen1
        \dimen1=0pt
        \ifdim  \shortenproofleft<0pt
        \then   \setbox\proofabove=\hbox{%
                        \kern-\shortenproofleft\unhbox\proofabove}%
                \shortenproofleft=0pt
        \fi
\fi
\ifdim  \dimen4<0pt
\then   \advance\shortenproofright\dimen4
        \advance\dimen0-\dimen4
        \dimen4=0pt
\fi
\ifdim  \shortenproofright<\wd\proofrulename
\then   \shortenproofright=\wd\proofrulename
\fi
\dimen2=\shortenproofleft \advance\dimen2 by\dimen1
\dimen3=\shortenproofright\advance\dimen3 by\dimen4
\ifproofdots
\then
        \dimen6=\shortenproofleft \advance\dimen6 .5\dimen0
        \setbox1=\vbox to\proofdotseparation{\vss\hbox{$\cdot$}\vss}%
        \setbox0=\hbox{%
                \advance\dimen6-.5\wd1
                \kern\dimen6
                $\vcenter to\proofdotnumber\proofdotseparation
                        {\leaders\box1\vfill}$%
                \unhbox\proofrulename}%
\else   \dimen6=\fontdimen22\the\textfont2 
        \dimen7=\dimen6
        \advance\dimen6by.5\proofrulebreadth
        \advance\dimen7by-.5\proofrulebreadth
        \setbox0=\hbox{%
                \kern\shortenproofleft
                \ifdoubleproof
                \then   \hbox to\dimen0{%
                        $\mathsurround0pt\mathord=\mkern-6mu%
                        \cleaders\hbox{$\mkern-2mu=\mkern-2mu$}\hfill
                        \mkern-6mu\mathord=$}%
                \else   \vrule height\dimen6 depth-\dimen7 width\dimen0
                \fi
                \unhbox\proofrulename}%
        \ht0=\dimen6 \dp0=-\dimen7
\fi
\let\doll\relax
\ifwasinsideprooftree
\then   \let\VBOX\vbox
\else   \ifmmode\else$\let\doll=$\fi
        \let\VBOX\vcenter
\fi
\VBOX   {\baselineskip\proofrulebaseline \lineskip.2ex
        \expandafter\lineskiplimit\ifproofdots0ex\else-0.6ex\fi
        \hbox   spread\dimen5   {\hfi\unhbox\proofabove\hfi}%
        \hbox{\box0}%
        \hbox   {\kern\dimen2 \box\proofbelow}}\doll%
\global\dimen2=\dimen2
\global\dimen3=\dimen3
\egroup 
\ifonleftofproofrule
\then   \shortenproofleft=\dimen2
\fi
\shortenproofright=\dimen3
\onleftofproofrulefalse
\ifinsideprooftree
\then   \hskip.5em plus 1fil \penalty2
\fi
}

\newcommand{\amr}[1]{\textbf{Comment by Amr:} #1 \textbf{End comment}}
\newcommand{\jjg}[1]{\textbf{Comment by Jon:} #1 \textbf{End comment}}

\newcommand{\omitnow}[1]{}

\newcommand{\ifo}{\mathbf{if}^{\circ}}

\newcommand{\ba}{\[\begin{array}{rcl}}
\newcommand{\ea}{\end{array}\]}
\newcommand{\Unit}{{\cal Q}_1}
\newcommand{\Qubit}{{\cal Q}_2}
\newcommand{\ie}{\textit{i.e.}}

\newcommand{\nf}{\textrm{nf}}

\newcommand{\varR}{\Ax{x \tin \sigma \vdash x \tin \sigma}{var}}

\newcommand{\letR}{
  \Ru{\G \vdash t \tin \sigma \qquad
      \D,x\tin\sigma \vdash u \tin \tau}
     {\G\ot\D \vdash \rlet x = t \,\rin\, u \tin \tau}
    {\rlet}
}

\newcommand{\unitR}{
  \Ru{\G,x:\Unit \vdash t \tin \sigma}
     {\G \vdash t \tin \sigma}
     {\mbox{wk-unit}}
}

\newcommand{\otintroR}{
  \Ru{\G \vdash t \tin \sigma \quad \D \vdash u \tin \tau}
     {\G \ot \D \vdash (t,u) \tin \sigma \ot \tau}
     {\ot\mbox{-intro}}
}

\newcommand{\otelimR}{
  \Ru{\G \vdash t \tin \sigma \ot \tau \qquad
       \D,\, x \tin \sigma, y \tin \tau \vdash u \tin \rho}
     {\G\ot\D \vdash \rlet (x,y) = t \rin u \tin \rho}
     {\ot\mbox{-elim}}
}

\newcommand{\fintrowR}{
  \Ax{\emptyC \vdash \mathrm{false} \tin \Qubit}
     {\mbox{f-intro}}
}

\newcommand{\tintrowR}{
  \Ax{\emptyC \vdash \mathrm{true} \tin \Qubit}
     {\mbox{t-intro}}
}

\newcommand{\ifoR}{
  \Ru{\begin{array}{l}
        \G \vdash c \tin \Qubit \qquad
        \D \vdash t,u \tin \sigma
    \end{array}}
     {\G\ot\D \vdash \rifo c \rthen t \relse u \tin \sigma}
    {\rifo}
}

\newcommand{\ifoRo}{
  \Ru{\begin{array}{l}
        \G \vdasho c \tin \Qubit \qquad
        \D \vdasho t,u \tin \sigma
 \qquad t \perp u
    \end{array}}
     {\G\ot\D \vdasho \rifo c \rthen t \relse u \tin \sigma}
    {\rifo}
}

\newcommand{\zeroR}{
  \Ax{\emptyC \vdash \overrightarrow{0} \tin \sigma}
   {\mbox{z-intro}}
}

\newcommand{\mulR}{
  \Ru{\Gamma \vdash t \tin \sigma}
     {\Gamma \vdash \kappa * t \tin \sigma}
   {\mbox{prob}}
}

\newcommand{\supR}{
 \Ru{\G\vdash t,u :\sigma} 
    {\G \vdash t+u \tin \sigma}
    {\mbox{sup}}
}

\newcommand{\supRo}{
 \Ru{\G\vdasho t,u :\sigma \qquad t \perp u \qquad |\lambda|^2+|\kappa|^2 = 1}
    {\G \vdasho \lambda*t+\kappa*u \tin \sigma}
    {\mbox{$\mathrm{sup}^\circ$}}
}

\newcommand{\substR}{
  \Ru{\G\vdasho t:\sigma\qquad \G \vdash t\equiv u : \sigma}
     {\G\vdasho u:\sigma}
     {\mbox{subst}}
}

\def\lastname{Altenkirch, Grattage, Vizzotto, and Sabry}

\begin{document}
\begin{frontmatter}
\title{An Algebra of Pure Quantum Programming}
  \author{Thorsten Altenkirch${}^{1}$ \qquad Jonathan Grattage\thanksref{txajonemail}}
  \address{The University of Nottingham, UK}

  \author{Juliana K. Vizzotto\thanksref{julianaemail}}
  \address{Federal University of Rio Grande do Sul, Brazil}

  \author{Amr Sabry\thanksref{amremail}}
  \address{Indiana University, USA}

  \thanks[txajonemail]{Email:
    {\texttt{\normalshape \{txa,jjg\}@cs.nottingham.ac.uk}}}
  \thanks[julianaemail]{Email: \href{mailto:jkv@inf.ufrgs.br}
    {\texttt{\normalshape jkv@inf.ufrgs.br}}}
  \thanks[amremail]{Email: \href{mailto:sabry@indiana.edu}
    {\texttt{\normalshape sabry@indiana.edu}}}

\begin{abstract}
We develop a sound and complete equational theory for the functional
quantum programming language QML. The soundness and completeness of
the theory are with respect to the previously-developed denotational
semantics of QML. The completeness proof also gives rise to a
normalisation algorithm following the \emph{normalisation by
evaluation} approach. The current work focuses on the pure fragment of
QML omitting measurements.
\end{abstract}

\begin{keyword}
quantum programming, completeness, normalisation
\end{keyword}

\end{frontmatter}

\section{Introduction}

The language QML was previously introduced by the first two
authors~\cite{alti:qml-draft}. Its semantics is inspired by the
denotational semantics of classical reversible computations. This
previous work provides a semantic foundation for reasoning about
quantum programs by mapping them to their denotations.

The natural next step is to develop reasoning principles on QML
programs themselves which avoid the detour via the denotational
semantics. For example, given the following QML definition of the
Hadamard gate:
\begingroup\par\noindent\advance\leftskip\mathindent\(
\begin{pboxed}\SaveRestoreHook
\column{B}{@{}l@{}}
\column{9}{@{}c@{}}
\column{9E}{@{}l@{}}
\column{12}{@{}l@{}}
\column{18}{@{}l@{}}
\column{28}{@{}l@{}}
\column{E}{@{}l@{}}
\fromto{B}{9}{{}\Conid{H}\;\Varid{x}{}}
\fromto{9}{9E}{{}\mathrel{=}{}}
\fromto{12}{E}{{}\mathbf{if}^\circ\;\Varid{x}{}}
\nextline
\fromto{12}{18}{{}\mathbf{then}\;{}}
\fromto{18}{28}{{}(\Varid{false}{}}
\fromto{28}{E}{{}\mathbin{+}(\mathbin{-}\mathrm{1})\mathbin{*}\Varid{true}){}}
\nextline
\fromto{12}{18}{{}\mathbf{else}\;{}}
\fromto{18}{28}{{}(\Varid{false}{}}
\fromto{28}{E}{{}\mathbin{+}\Varid{true}){}}
\ColumnHook
\end{pboxed}
\)\par\noindent\endgroup\resethooks
We would like to verify that \ensuremath{\Conid{H}\;(\Conid{H}\;\Varid{x})} is observationally equivalent
to \ensuremath{\Varid{x}}, using a derivation like:

\begingroup\par\noindent\advance\leftskip\mathindent\(
\begin{pboxed}\SaveRestoreHook
\column{B}{@{}l@{}}
\column{9}{@{}c@{}}
\column{9E}{@{}l@{}}
\column{12}{@{}l@{}}
\column{17}{@{}c@{}}
\column{17E}{@{}l@{}}
\column{18}{@{}l@{}}
\column{20}{@{}l@{}}
\column{24}{@{}l@{}}
\column{26}{@{}l@{}}
\column{E}{@{}l@{}}
\fromto{B}{9}{{}\Conid{H}\;(\Conid{H}\;\Varid{x}){}}
\fromto{9}{9E}{{}\mathrel{=}{}}
\fromto{12}{17}{{}\mathbf{if}^\circ\;{}}
\fromto{17}{17E}{{}({}}
\fromto{20}{E}{{}\mathbf{if}^\circ\;\Varid{x}{}}
\nextline
\fromto{20}{26}{{}\mathbf{then}\;{}}
\fromto{26}{E}{{}(\Varid{false}\mathbin{+}(\mathbin{-}\mathrm{1})\mathbin{*}\Varid{true}){}}
\nextline
\fromto{20}{26}{{}\mathbf{else}\;{}}
\fromto{26}{E}{{}(\Varid{false}\mathbin{+}\Varid{true})){}}
\nextline
\fromto{12}{18}{{}\mathbf{then}\;{}}
\fromto{18}{E}{{}(\Varid{false}\mathbin{+}(\mathbin{-}\mathrm{1})\mathbin{*}\Varid{true}){}}
\nextline
\fromto{12}{18}{{}\mathbf{else}\;{}}
\fromto{18}{E}{{}(\Varid{false}\mathbin{+}\Varid{true}){}}
\nextline[\blanklineskip]
\fromto{12}{E}{{}\mbox{\onelinecomment by commuting conversion for \ensuremath{\mathbf{if}^\circ}}{}}
\nextline[\blanklineskip]
\fromto{9}{9E}{{}\mathrel{=}{}}
\fromto{12}{E}{{}\mathbf{if}^\circ\;\Varid{x}{}}
\nextline
\fromto{12}{18}{{}\mathbf{then}\;{}}
\fromto{18}{E}{{}\mathbf{if}^\circ\;(\Varid{false}\mathbin{+}(\mathbin{-}\mathrm{1})\mathbin{*}\Varid{true}){}}
\nextline
\fromto{18}{24}{{}\mathbf{then}\;{}}
\fromto{24}{E}{{}(\Varid{false}\mathbin{+}(\mathbin{-}\mathrm{1})\mathbin{*}\Varid{true}){}}
\nextline
\fromto{18}{24}{{}\mathbf{else}\;{}}
\fromto{24}{E}{{}(\Varid{false}\mathbin{+}\Varid{true}){}}
\nextline
\fromto{12}{18}{{}\mathbf{else}\;{}}
\fromto{18}{E}{{}\mathbf{if}^\circ\;(\Varid{false}\mathbin{+}\Varid{true}){}}
\nextline
\fromto{18}{24}{{}\mathbf{then}\;{}}
\fromto{24}{E}{{}(\Varid{false}\mathbin{+}(\mathbin{-}\mathrm{1})\mathbin{*}\Varid{true}){}}
\nextline
\fromto{18}{24}{{}\mathbf{else}\;{}}
\fromto{24}{E}{{}(\Varid{false}\mathbin{+}\Varid{true}){}}
\nextline[\blanklineskip]
\fromto{12}{E}{{}\mbox{\onelinecomment by $\ifo$}{}}
\nextline[\blanklineskip]
\fromto{9}{9E}{{}\mathrel{=}{}}
\fromto{12}{E}{{}\mathbf{if}^\circ\;\Varid{x}{}}
\nextline
\fromto{12}{18}{{}\mathbf{then}\;{}}
\fromto{18}{E}{{}(\Varid{false}\mathbin{-}\Varid{false}\mathbin{+}\Varid{true}\mathbin{+}\Varid{true}){}}
\nextline
\fromto{12}{18}{{}\mathbf{else}\;{}}
\fromto{18}{E}{{}(\Varid{false}\mathbin{+}\Varid{false}\mathbin{+}\Varid{true}\mathbin{-}\Varid{true}){}}
\nextline[\blanklineskip]
\fromto{12}{E}{{}\mbox{\onelinecomment by simplification and normalisation}{}}
\nextline[\blanklineskip]
\fromto{9}{9E}{{}\mathrel{=}{}}
\fromto{12}{E}{{}\mathbf{if}^\circ\;\Varid{x}\;\mathbf{then}\;\Varid{true}\;\mathbf{else}\;\Varid{false}{}}
\nextline[\blanklineskip]
\fromto{12}{E}{{}\mbox{\onelinecomment by $\eta$-rule for $\ifo$}{}}
\nextline[\blanklineskip]
\fromto{9}{9E}{{}\mathrel{=}{}}
\fromto{12}{E}{{}\Varid{x}{}}
\ColumnHook
\end{pboxed}
\)\par\noindent\endgroup\resethooks

It is relatively easy to develop \emph{some} set of sound equational
principles. Inspired by equivalences on classical computations, one
may hypothesise that certain equations should hold and simply verify
that both sides of the equation have the same denotation.

Given, however, that QML is based on a first-order functional language
with finite types, it should be possible to also develop a
\emph{complete} set of equivalences that totally capture denotational
equivalence. Technically, one can prove completeness of the equational
semantics by ``inverting'' the denotational meaning function. The
construction is subtle in parts. We present it first in the context
of the classical sublanguage of QML, and then extend it to deal with
quantum data and control.

The paper is thus organised as follows. We begin with an informal
review of QML in Section~\ref{sec:qml}. In Section~\ref{sec:classsem},
we present the denotational semantics of the classical sublanguage of
QML, and present a system of equations that is sound with respect to
the denotational semantics. We then show that this set of equations is
complete in Section~\ref{sec:compclass}. Section~\ref{sec:quantum}
repeats the development for the quantum
constructs. Section~\ref{sec:conc} concludes.

\omitnow{
The appendices include proofs for some of the main lemmas, and
concrete definitions of the categorical semantics using Haskell as a
meta-language.
}

\section{Related work}
\label{sec:related}

Peter Selinger's influential paper \cite{selinger:qpl} introduces a
single-assignment (essentially functional) quantum programming
language, which is based on the separation of \emph{classical control}
and \emph{quantum data}. This language combines high-level classical
structures with operations on quantum data, and has a clear
mathematical semantics in the form of superoperators. Quantum data can
be manipulated by using unitary operators or by measurement, which can
effect the classical control flow.

Recently, Selinger and Valiron \cite{selingerValiron:lambda} have
presented a functional language based on the same \emph{classical
control} and \emph{quantum data} paradigm. Selinger and Valiron's
approach is in some sense complementary to ours: they use an affine
type system (no contraction), while we use a strict system (no
weakening). The lack of contraction is justified by the no-cloning
property of quantum states. However, this does not apply to our
approach, since we model contraction by sharing not by copying ---
this is also used in the calculus of Arrighi and
Dowek~\cite{arrighiDowek}.

Andre van Tonder~\cite{tonder1,tonder2} has proposed a quantum
$\lambda$-calculus incorporating higher order programs, but no
measurements. He also suggests an equational theory for strict (higher
order) computations, but shows neither completeness nor normalisation.

\section{QML Syntax and Examples}
\label{sec:qml}

The QML terms consist of those of a first-order functional language, extended
with quantum data and quantum control. The full language also includes
quantum measurement, which we do not consider in this paper. The syntax of
terms is the following:

\begingroup\par\noindent\advance\leftskip\mathindent\(
\begin{pboxed}\SaveRestoreHook
\column{B}{@{}l@{}}
\column{21}{@{}l@{}}
\column{39}{@{}l@{}}
\column{44}{@{}l@{}}
\column{E}{@{}l@{}}
\fromto{B}{21}{{}(\Conid{Variables})\;{}}
\fromto{21}{39}{{}\Varid{x},\Varid{y},\mathbin{...}{}}
\fromto{39}{44}{{}\in\;{}}
\fromto{44}{E}{{}\Conid{Vars}{}}
\nextline[\blanklineskip]
\fromto{B}{21}{{}(\Conid{Prob}.\Varid{amplitudes})\;{}}
\fromto{21}{39}{{}\kappa,\iota,\mathbin{...}{}}
\fromto{39}{44}{{}\in\;{}}
\fromto{44}{E}{{}\mathbb{C}{}}
\nextline
\fromto{B}{21}{{}(\Conid{Patterns})\;{}}
\fromto{21}{39}{{}\Varid{p},\Varid{q}{}}
\fromto{39}{44}{{}\mathbin{::=}{}}
\fromto{44}{E}{{}\Varid{x}\mid (\Varid{x},\Varid{y}){}}
\nextline
\fromto{B}{21}{{}(\Conid{Terms})\;{}}
\fromto{21}{39}{{}\Varid{t},\Varid{u},\Varid{e}{}}
\fromto{39}{44}{{}\mathbin{::=}{}}
\fromto{44}{E}{{}\Varid{x}\mid ()\mid (\Varid{t},\Varid{u}){}}
\nextline
\fromto{39}{44}{{}\mid {}}
\fromto{44}{E}{{}\mathbf{let}\;\Varid{p}\mathrel{=}\Varid{t}\;\mathbf{in}\;\Varid{u}{}}
\nextline
\fromto{39}{44}{{}\mid {}}
\fromto{44}{E}{{}\mathbf{if}^\circ\;\Varid{t}\;\mathbf{then}\;\Varid{u}\;\mathbf{else}\;\Varid{u'}{}}
\nextline
\fromto{39}{44}{{}\mid {}}
\fromto{44}{E}{{}\Varid{false}\mid \Varid{true}\mid \overrightarrow{0}\mid \kappa\mathbin{*}\Varid{t}\mid \Varid{t}\mathbin{+}\Varid{u}{}}
\ColumnHook
\end{pboxed}
\)\par\noindent\endgroup\resethooks

The classic sublanguage consists of variables, \ensuremath{\mathbf{let}}-expressions,
unit, pairs, booleans, and conditionals. Quantum data is modelled
using the constructs \ensuremath{\kappa\mathbin{*}\Varid{t}}, \ensuremath{\overrightarrow{0}}, and \ensuremath{\Varid{t}\mathbin{+}\Varid{u}}. The term \ensuremath{\kappa\mathbin{*}\Varid{t}} where \ensuremath{\kappa} is a complex number associates the \emph{probability
amplitude} \ensuremath{\kappa} with the term \ensuremath{\Varid{t}}. It is convenient to have a
special constant \ensuremath{\overrightarrow{0}} for terms with probability amplitude zero. The
term \ensuremath{\Varid{t}\mathbin{+}\Varid{u}} is a quantum \emph{superposition} of \ensuremath{\Varid{t}} and \ensuremath{\Varid{u}}. Quantum
superpositions are first-class values: when used as the first
subexpression of a conditional, they turn the conditional into a
\emph{quantum control} construct. For example, \ensuremath{\mathbf{if}^\circ\;(\Varid{true}\mathbin{+}\Varid{false})\;\mathbf{then}\;\Varid{t}\;\mathbf{else}\;\Varid{u}} evaluates both \ensuremath{\Varid{t}} and \ensuremath{\Varid{u}} and combines their results
in a quantum superposition.

\omitnow{
Quantum data includes \emph{superpositions} \ensuremath{\{\mskip1.5mu (\kappa)\;\Varid{false}\mid (\iota)\;\Varid{true}\mskip1.5mu\}} of the basic boolean observables.  The term \ensuremath{\{\mskip1.5mu (\mathrm{1}\mathbin{/}\sqrt{\mathrm{2}})\;\Varid{false}\mid (\mathrm{1}\mathbin{/}\sqrt{\mathrm{2}})\;\Varid{true}\mskip1.5mu\}} is an equal superposition of \ensuremath{\Varid{false}} and
\ensuremath{\Varid{true}}. We sometimes omit the normalisation factors, which can be
inferred, and write \ensuremath{\{\mskip1.5mu \Varid{false}\mid \Varid{true}\mskip1.5mu\}} instead.  If one of the
coefficients is \ensuremath{\mathrm{0}} it may also be omitted, \emph{e.g.,} we write
\ensuremath{\Varid{true}} for the qubit presenting the classical value \ensuremath{\Varid{true}} and
\ensuremath{\{\mskip1.5mu (\mbox{$i$})\;\Varid{true}\mskip1.5mu\}} to construct a qubit which behaves like \ensuremath{\Varid{true}} if
measured, but which has a different phase.
}

\subsection{Examples}

To give further intuition about the semantics of QML, we consider a
few more interesting examples. In the examples, we allow the
definition and use of ``global'' function symbols. Adding such
definitions to the formalism is possible but tedious, so we keep them
at an informal meta-level.

The following three functions correspond to simple rotations on
qubits:
\begingroup\par\noindent\advance\leftskip\mathindent\(
\begin{pboxed}\SaveRestoreHook
\column{B}{@{}l@{}}
\column{3}{@{}l@{}}
\column{11}{@{}c@{}}
\column{11E}{@{}l@{}}
\column{14}{@{}l@{}}
\column{21}{@{}l@{}}
\column{26}{@{}l@{}}
\column{27}{@{}l@{}}
\column{38}{@{}l@{}}
\column{50}{@{}l@{}}
\column{56}{@{}l@{}}
\column{E}{@{}l@{}}
\fromto{3}{11}{{}\Varid{qnot}\;\Varid{x}{}}
\fromto{11}{11E}{{}\mathrel{=}{}}
\fromto{14}{26}{{}\mathbf{if}^\circ\;\Varid{x}\;\mathbf{then}\;{}}
\fromto{26}{38}{{}\Varid{false}\;\mathbf{else}\;{}}
\fromto{38}{E}{{}\Varid{true}{}}
\nextline[\blanklineskip]
\fromto{3}{11}{{}\Varid{had}\;\Varid{x}{}}
\fromto{11}{11E}{{}\mathrel{=}{}}
\fromto{14}{21}{{}\mathbf{if}^\circ\;\Varid{x}\;{}}
\fromto{21}{27}{{}\mathbf{then}\;{}}
\fromto{27}{50}{{}((\mathbin{-}\mathrm{1})\mathbin{*}\Varid{true}\mathbin{+}\Varid{false})\;{}}
\fromto{50}{56}{{}\mathbf{else}\;{}}
\fromto{56}{E}{{}(\Varid{true}\mathbin{+}\Varid{false}){}}
\nextline[\blanklineskip]
\fromto{3}{11}{{}\Varid{z}\;\Varid{x}{}}
\fromto{11}{11E}{{}\mathrel{=}{}}
\fromto{14}{21}{{}\mathbf{if}^\circ\;\Varid{x}\;{}}
\fromto{21}{E}{{}\mathbf{then}\;(\mbox{$i$}\mathbin{*}\Varid{true})\;\mathbf{else}\;\Varid{false}{}}
\ColumnHook
\end{pboxed}
\)\par\noindent\endgroup\resethooks
The first is the quantum version of boolean negation: it behaves as
usual when applied to classical values but it also applies to quantum
data. Evaluating \ensuremath{\Varid{qnot}\;(\kappa\mathbin{*}\Varid{false}\mathbin{+}\iota\mathbin{*}\Varid{true})} swaps
the probability amplitudes associated with \ensuremath{\Varid{false}} and \ensuremath{\Varid{true}}. The
second function represents the fundamental \emph{Hadamard} matrix, and
the third represents the \emph{phase} gate.

The function:
\begingroup\par\noindent\advance\leftskip\mathindent\(
\begin{pboxed}\SaveRestoreHook
\column{B}{@{}l@{}}
\column{3}{@{}l@{}}
\column{15}{@{}l@{}}
\column{21}{@{}l@{}}
\column{30}{@{}l@{}}
\column{E}{@{}l@{}}
\fromto{3}{15}{{}cnot \;\Varid{c}\;\Varid{x}\mathrel{=}{}}
\fromto{15}{E}{{}\mathbf{if}^\circ\;\Varid{c}{}}
\nextline
\fromto{15}{21}{{}\mathbf{then}\;{}}
\fromto{21}{30}{{}(\Varid{true},{}}
\fromto{30}{E}{{}\Varid{qnot}\;\Varid{x}){}}
\nextline
\fromto{15}{21}{{}\mathbf{else}\;{}}
\fromto{21}{30}{{}(\Varid{false},{}}
\fromto{30}{E}{{}\Varid{x}){}}
\ColumnHook
\end{pboxed}
\)\par\noindent\endgroup\resethooks
is the conditional-not operation, which behaves as follows: if the
control qubit~\ensuremath{\Varid{c}} is \ensuremath{\Varid{true}} it negates the second qubit \ensuremath{\Varid{x}}; otherwise
it leaves it unchanged. When the control qubit is in some
superposition of \ensuremath{\Varid{true}} and \ensuremath{\Varid{false}}, the result is a superposition of
the two pairs resulting from the evaluation of each branch of the
conditional. For example, evaluating \ensuremath{cnot \;(\Varid{false}\mathbin{+}\Varid{true})\;\Varid{false}}
produces the \emph{entangled} pair \ensuremath{(\Varid{false},\Varid{false})\mathbin{+}(\Varid{true},\Varid{true})}.

\subsection{Copying and Discarding Quantum Data}

To motivate the main aspects of the type system in the next section, we
examine in detail the issues related to copying and discarding quantum
data.

A simple example where quantum data appears to be copied, in violation of the
\emph{no-cloning} theorem~\cite{NC00}, is:
\begingroup\par\noindent\advance\leftskip\mathindent\(
\begin{pboxed}\SaveRestoreHook
\column{B}{@{}l@{}}
\column{E}{@{}l@{}}
\fromto{B}{E}{{}\mathbf{let}\;\Varid{x}\mathrel{=}\Varid{false}\mathbin{+}\Varid{true}{}}
\nextline
\fromto{B}{E}{{}\mathbf{in}\;(\Varid{x},\Varid{x}){}}
\ColumnHook
\end{pboxed}
\)\par\noindent\endgroup\resethooks
As the formal semantics of QML clarifies, this expression does not actually
clone quantum data; rather it \emph{shares} one copy of the quantum data.
With this interpretation, one can freely duplicate variables bound to quantum
data. When translated to the type system, this means that the type system
imposes no restrictions on the use of the structural rule of
\emph{contraction}.

Discarding variables bound to quantum data is however
problematic. Consider the expression:
\begingroup\par\noindent\advance\leftskip\mathindent\(
\begin{pboxed}\SaveRestoreHook
\column{B}{@{}l@{}}
\column{E}{@{}l@{}}
\fromto{B}{E}{{}\mathbf{let}\;(\Varid{x},\Varid{y})\mathrel{=}(\Varid{false},\Varid{false})\mathbin{+}(\Varid{true},\Varid{true}){}}
\nextline
\fromto{B}{E}{{}\mathbf{in}\;\Varid{x}{}}
\ColumnHook
\end{pboxed}
\)\par\noindent\endgroup\resethooks
where the quantum data bound to \ensuremath{\Varid{y}} is discarded.  According to both
the physical interpretations of quantum computation, and the semantics
of QML, this corresponds to a \emph{measurement} of \ensuremath{\Varid{y}}. Since
measurement is semantically quite complicated to deal with, we insist
that it should be represented explicitly. The language we consider in
this paper lacks the explicit constructs for measurement so we reject
the expression above. This means that the structural rule of
\emph{weakening} is never allowed in situations where information may
be lost.

\section{The Classical Sublanguage}
\label{sec:classsem}
\label{sec:classcomplete}

By the classical sublanguage, we mean the subset of terms excluding quantum
superpositions and hence quantum control.



\subsection{Type System}

The main r\^{o}le of the type system is to control the use of variables. The
typing rules of QML are based on strict linear logic, where contractions are
implicit and weakenings are not allowed when they correspond to information
loss. As explained in the previous section, weakenings correspond to
measurements, which are not supported in the subset of the language discussed
in this paper.

We use \ensuremath{\sigma,\tau,\rho} to vary over QML types which are given by
the following grammar:
\begingroup\par\noindent\advance\leftskip\mathindent\(
\begin{pboxed}\SaveRestoreHook
\column{B}{@{}l@{}}
\column{8}{@{}l@{}}
\column{E}{@{}l@{}}
\fromto{B}{8}{{}\sigma{}}
\fromto{8}{E}{{}\mathrel{=}\Unit\mid \Qubit\mid \sigma\otimes\tau{}}
\ColumnHook
\end{pboxed}
\)\par\noindent\endgroup\resethooks
As apparent from the grammar, QML types are first-order and finite: there
are no higher-order types and no recursive types. The only types we can
represent are the types of collections of qubits.

Typing contexts (\ensuremath{\Gamma,\Delta}) are given by:
\begingroup\par\noindent\advance\leftskip\mathindent\(
\begin{pboxed}\SaveRestoreHook
\column{B}{@{}l@{}}
\column{E}{@{}l@{}}
\fromto{B}{E}{{}\Gamma\mathrel{=}\bullet\mid \Gamma,\Varid{x}\mathbin{:}\sigma{}}
\ColumnHook
\end{pboxed}
\)\par\noindent\endgroup\resethooks
where \ensuremath{\bullet} stands for the empty context, but is omitted if the context is
non-empty. For simplicity we assume that every variable appears at most
once. Contexts correspond to functions from a finite set of variables to
types. We introduce the operator $\otimes$, mapping pairs of contexts to
contexts:
\[\begin{array}{lcll}
  (\Gamma,x:\sigma) \otimes (\Delta,x:\sigma) & = &
  (\Gamma\otimes\Delta),x:\sigma\\
  (\Gamma,x:\sigma) \otimes \Delta & = & (\Gamma \otimes
  \Delta),x:\sigma & \mbox{if $x \notin \dom{(\Delta)}$}\\
  \emptyC \otimes \Delta & = & \Delta
\end{array}\]
This operation is partial: it is only well-defined if the two contexts
do not assign different types to the same variable. Whenever we use
this operator we implicitly assume that it is well-defined.


Figure~\ref{fig:typingterms} presents the rules for deriving valid typing
judgements $\Gamma\vdash t:\sigma$. 
The only variables that may be dropped from the context are the ones
of type~$\Unit$ which, by definition, carry no information. Otherwise
the type system forces every variable in the context to be used
(perhaps more than once if it is shared).

\begin{figure*}[t]
\centering{\small
\framebox{
$\begin{array}{c}
\varR \qquad \letR \\
\Ax{\emptyC \vdash () \tin \Unit}{unit} \qquad \otintroR \\
\otelimR \\
\fintrowR \qquad \tintrowR \\
\ifoR \qquad \unitR 
\end{array}$}}
\caption{Typing classical terms}
\label{fig:typingterms}
\end{figure*}

\subsection{The Category of Typed Terms}

The set of typed terms can be organised in an elegant categorical structure,
which facilitates the proofs later. The objects of the category are contexts;
the homset between the objects $\G$ and $\D$, denoted $\Tmm{\G}{\D}$,
consists of all the terms $t$ such that $\G\vdash t:\con{\Delta}$ where
$\con{\Delta}$ views the context $\Delta$ as a type.  This latter map is
naturally defined as follows:

\ba
\con{\emptyC} &=& \Unit \\
\con{\Gamma,x:\sigma} &=& \con{\G} \ot {\sigma} 
\ea



For each context $\G$, the identity $1_\G \in \Tm{\G}{\G}$ is defined as follows:
\ba
1_\emptyC & = & () \\
1_{\G,x:\sigma} & = & (1_\G,x)
\ea


To express composition, we first define:
\begingroup\par\noindent\advance\leftskip\mathindent\(
\begin{pboxed}\SaveRestoreHook
\column{B}{@{}l@{}}
\column{3}{@{}l@{}}
\column{32}{@{}c@{}}
\column{32E}{@{}l@{}}
\column{37}{@{}l@{}}
\column{E}{@{}l@{}}
\fromto{3}{32}{{}\textbf{let}^\mathbf{*}\;\bullet\mathrel{=}\Varid{u}\;\mathbf{in}\;\Varid{t}{}}
\fromto{32}{32E}{{}\qquad\equiv\qquad{}}
\fromto{37}{E}{{}\Varid{t}{}}
\nextline
\fromto{3}{32}{{}\textbf{let}^\mathbf{*}\;\Gamma,\Varid{x}\mathbin{:}\sigma\mathrel{=}\Varid{u}\;\mathbf{in}\;\Varid{t}{}}
\fromto{32}{32E}{{}\qquad\equiv\qquad{}}
\fromto{37}{E}{{}\mathbf{let}\;(\Varid{x}_{\Varid{r}},\Varid{x})\mathrel{=}\Varid{u}\;\mathbf{in}\;\textbf{let}^\mathbf{*}\;\Gamma\mathrel{=}\Varid{x}_{\Varid{r}}\;\mathbf{in}\;\Varid{t}{}}
\ColumnHook
\end{pboxed}
\)\par\noindent\endgroup\resethooks
Given $d\in\Tmm{\D}{\G}$ and $e\in\Tmm{\G}{\Theta}$, the composition $e \circ
d\in\Tmm{\D}{\Theta}$ is given by the term $\ensuremath{\textbf{let}^\mathbf{*}\;\Gamma\mathrel{=}\Varid{d}\;\mathbf{in}\;\Varid{e}}$.


\subsection{Semantics}

The intention is to interpret every type $\sigma$ and every context $\Gamma$
as finite sets $\evalC{\sigma}$ and $\evalC{\Gamma}$, and then interpret a
judgement $\Gamma \vdash t \tin \sigma$ as a function 
$\evalC{\G\vdash t\tin\sigma}\in\evalC{\Gamma}\to\evalC{\sigma}$.

In the classical case, the type $\Qubit$ is simply the type of booleans; the types
are interpreted as follows:
\ba
\evalC{\Unit} &=& \{ 0 \} \\
\evalC{\Qubit} &=& \{ 0,1 \} \\
\evalC{\sigma \ot \tau} &=& \evalC{\sigma} \times \evalC{\tau} \ea
We use the abbreviation $\evalC{\G}$ for $\evalC{\con{\G}}$.


The meaning function is defined in Figure~\ref{fig:classmean} by
induction over the structure of type derivations. It uses the
following auxiliary maps:
\begin{itemize}
\item $\mathit{id} : S \to S$ defined by $\mathit{id}(a) = a$
\item $\mathit{id}^* : S \to \evalC{\Unit}\times S$ and its inverse
$\mathit{id}_*$ defined by $\mathit{id^*}(a) = (0,a)$ and
$\mathit{id_*}(0,a) = a$
\item For $a \in S$, the family of constant functions $\vconst{a} :
  \evalC{\Unit} \to S$ defined by
  $(\vconst{a})(0)=a$.
\item $\delta : S \to (S,S)$ defined by $\delta(a) = (a,a)$
\item $\mathit{swap} : S \times T \to T \times S$ defined by
$\mathit{swap}(a,b) = (b,a)$. We will usually implicitly use
$\textit{swap}$ to avoid cluttering the figures with maps which just
re-shuffle values.
\item For any two functions $f \in S_1 \to T_1$ and $g \in S_2 \to
T_2$, the function $(f \times g) : (S_1 \times S_2) \to (T_1 \times
T_2)$ is defined as usual:
\ba
(f \times g) (a,b) &=& (f~a, g~b)
\ea
\item $\delta_{\G,\D} : \evalC{\Gamma\ot\Delta} \to
\evalC{\Gamma}\times\evalC{\Delta}$. This map is defined by induction
on the definition of $\Gamma\ot\Delta$ as follows:
\[
\delta_{\G,\D} =
 \left\{ \begin{array}{rl}
  \delta_{\G',\D'} \times \delta &
    \mbox{if~} \G=\G',x:\sigma \mbox{~and~}
               \D=\D',x:\sigma \\
  \delta_{\G',\D} \times \mathit{id} &
    \mbox{if~} \G=\G',x:\sigma \mbox{~and~}
    x \not\in \dom{(\Delta)} \\
  \mathit{id}^* & \mbox{if~} \G=\emptyC
   \end{array}\right.
\]
Intuitively, the map $\delta_{\G,\D}$ takes an incoming environment for an
expression, creates shared copies of the appropriate values, and rearranges
them (the shuffling is implicit and not shown in the above definition) into
two environments that are then passed to the subexpressions.
\item For any two functions $f,g \in S \to T$, we define
the conditional $\cond{f}{g} \in (\evalC{\Qubit} \times S) \to T$ as follows:
\ba
(\cond{f}{g})~(1,a) &=& f~a \\
(\cond{f}{g})~(0,a) &=& g~a
\ea
\end{itemize}

\begin{figure*}[t]
\centering{\small
\framebox{
$\begin{array}{rcl}
\evalC{{\emptyC \vdash () \tin \Unit}} &=& \vconst{0}
\\
\evalC{{\emptyC \vdash \mathrm{false} \tin \Qubit}} &=& \vconst{0}
\\
\evalC{{\emptyC \vdash \mathrm{true} \tin \Qubit}} &=&  \vconst{1}
\\
\evalC{{x:\sigma \vdash x \tin \sigma}} &=& \mathit{id}_*
\\
\evalC{{\G\ot\D \vdash \rlet x = t \,\rin\, u \tin \tau}} &=&
   g \circ (f \times \mathit{id}) \circ \delta_{\G,\D} \\
    && \mbox{~where~} \begin{array}[t]{rcl}
    f &=& \evalC{{\G \vdash t \tin \sigma}} \\
    g &=& \evalC{{\D,x\tin\sigma \vdash u \tin \tau}}
  \end{array}
\\
\evalC{{\G \ot \D \vdash (t,u) \tin \sigma \ot \tau}} &=&
   (f \times g) \circ \delta_{\G,\D} \\
   && \mbox{~where~} \begin{array}[t]{rcl}
    f &=& \evalC{{\G \vdash t \tin \sigma}} \\
    g &=& \evalC{{\D \vdash u \tin \tau}}
   \end{array}
\\
\evalC{{\G\ot\D \vdash \rlet (x,y) = t \rin u \tin \rho}} &=&
   g \circ (f \times \mathit{id}) \circ \delta_{\G,\D} \\
   && \mbox{~where~} \begin{array}[t]{rcl}
    f &=& \evalC{{\G \vdash t \tin \sigma \ot \tau}} \\
    g &=& \evalC{{\D,\, x \tin \sigma, y \tin \tau \vdash u \tin \rho}}
   \end{array}
\\
\evalC{{\G\ot\D \vdash \rifo c \rthen t \relse u \tin \sigma}} &=&
  (\cond{g}{h}) \circ (f \times \mathit{id}) \circ \delta_{\G,\D} \\
   && \mbox{~where~} \begin{array}[t]{rcl}
    f &=& \evalC{{\G \vdash c \tin \Qubit}} \\
    g &=& \evalC{{\D \vdash t \tin \sigma}} \\
    h &=& \evalC{{\D \vdash u \tin \sigma}}
  \end{array}
\\
\evalC{\G\vdash t \tin\sigma} &=& f \circ \mathit{id}^* \\
   && \mbox{~where~} f = \evalC{\G,x:\Unit\vdash t \tin \sigma}
\end{array}$}}
\caption{Meaning of classical derivations}
\label{fig:classmean}
\end{figure*}

\subsection{Equational Theory}

We present the equational theory for the classical sublanguage and
then show its soundness and completeness. The equations refer to a set
of syntactic values defined as follows:
\begingroup\par\noindent\advance\leftskip\mathindent\(
\begin{pboxed}\SaveRestoreHook
\column{B}{@{}l@{}}
\column{E}{@{}l@{}}
\fromto{B}{E}{{}\Varid{val}\;\in\; \ValC \mathbin{::=}\Varid{x}\mid ()\mid \Varid{false}\mid \Varid{true}\mid (\Varid{val}_{\mathrm{1}},\Varid{val}_{\mathrm{2}}){}}
\ColumnHook
\end{pboxed}
\)\par\noindent\endgroup\resethooks

\begin{definition}
  The \emph{classical equations} are grouped in four categories.
\begin{itemize}
\item \ensuremath{\mathbf{let}}-equation
\begingroup\par\noindent\advance\leftskip\mathindent\(
\begin{pboxed}\SaveRestoreHook
\column{B}{@{}l@{}}
\column{26}{@{}c@{}}
\column{26E}{@{}l@{}}
\column{31}{@{}l@{}}
\column{E}{@{}l@{}}
\fromto{B}{26}{{}\mathbf{let}\;\Varid{p}\mathrel{=}\Varid{val}\;\mathbf{in}\;\Varid{u}{}}
\fromto{26}{26E}{{}\qquad\equiv\qquad{}}
\fromto{31}{E}{{}\Varid{u}\;[\mskip1.5mu \Varid{val}\mathbin{/}\Varid{p}\mskip1.5mu]{}}
\ColumnHook
\end{pboxed}
\)\par\noindent\endgroup\resethooks
\item $\beta$-equations
\begingroup\par\noindent\advance\leftskip\mathindent\(
\begin{pboxed}\SaveRestoreHook
\column{B}{@{}l@{}}
\column{11}{@{}l@{}}
\column{27}{@{}c@{}}
\column{27E}{@{}l@{}}
\column{32}{@{}l@{}}
\column{E}{@{}l@{}}
\fromto{B}{27}{{}\mathbf{let}\;(\Varid{x},\Varid{y})\mathrel{=}(\Varid{t},\Varid{u})\;\mathbf{in}\;\Varid{e}{}}
\fromto{27}{27E}{{}\qquad\equiv\qquad{}}
\fromto{32}{E}{{}\mathbf{let}\;\Varid{x}\mathrel{=}\Varid{t}\;\mathbf{in}\;\mathbf{let}\;\Varid{y}\mathrel{=}\Varid{u}\;\mathbf{in}\;\Varid{e}{}}
\nextline
\fromto{B}{27}{{}\mathbf{if}^\circ\;\Varid{false}\;\mathbf{then}\;\Varid{t}\;\mathbf{else}\;\Varid{u}{}}
\fromto{27}{27E}{{}\qquad\equiv\qquad{}}
\fromto{32}{E}{{}\Varid{u}{}}
\nextline
\fromto{B}{11}{{}\mathbf{if}^\circ\;\Varid{true}\;{}}
\fromto{11}{27}{{}\mathbf{then}\;\Varid{t}\;\mathbf{else}\;\Varid{u}{}}
\fromto{27}{27E}{{}\qquad\equiv\qquad{}}
\fromto{32}{E}{{}\Varid{t}{}}
\ColumnHook
\end{pboxed}
\)\par\noindent\endgroup\resethooks
\item $\eta$-equations
\begingroup\par\noindent\advance\leftskip\mathindent\(
\begin{pboxed}\SaveRestoreHook
\column{B}{@{}l@{}}
\column{30}{@{}c@{}}
\column{30E}{@{}l@{}}
\column{35}{@{}l@{}}
\column{38}{@{}l@{}}
\column{E}{@{}l@{}}
\fromto{B}{30}{{}(){}}
\fromto{30}{30E}{{}\qquad\equiv\qquad{}}
\fromto{35}{38}{{}\Varid{t}{}}
\fromto{38}{E}{{}\mbox{\onelinecomment  if t:\ensuremath{\Unit}}{}}
\nextline
\fromto{B}{30}{{}\mathbf{let}\;\Varid{x}\mathrel{=}\Varid{t}\;\mathbf{in}\;\Varid{x}{}}
\fromto{30}{30E}{{}\qquad\equiv\qquad{}}
\fromto{35}{E}{{}\Varid{t}{}}
\nextline
\fromto{B}{30}{{}\mathbf{let}\;(\Varid{x},\Varid{y})\mathrel{=}\Varid{t}\;\mathbf{in}\;(\Varid{x},\Varid{y}){}}
\fromto{30}{30E}{{}\qquad\equiv\qquad{}}
\fromto{35}{E}{{}\Varid{t}{}}
\nextline
\fromto{B}{30}{{}\mathbf{if}^\circ\;\Varid{t}\;\mathbf{then}\;\Varid{true}\;\mathbf{else}\;\Varid{false}{}}
\fromto{30}{30E}{{}\qquad\equiv\qquad{}}
\fromto{35}{E}{{}\Varid{t}{}}
\ColumnHook
\end{pboxed}
\)\par\noindent\endgroup\resethooks
\item Commuting conversions
\begingroup\par\noindent\advance\leftskip\mathindent\(
\begin{pboxed}\SaveRestoreHook
\column{B}{@{}l@{}}
\column{5}{@{}l@{}}
\column{14}{@{}l@{}}
\column{35}{@{}c@{}}
\column{35E}{@{}l@{}}
\column{40}{@{}l@{}}
\column{46}{@{}l@{}}
\column{E}{@{}l@{}}
\fromto{5}{35}{{}\mathbf{let}\;\Varid{p}\mathrel{=}\Varid{t}\;\mathbf{in}\;\mathbf{let}\;\Varid{q}\mathrel{=}\Varid{u}\;\mathbf{in}\;\Varid{e}{}}
\fromto{35}{35E}{{}\qquad\equiv\qquad{}}
\fromto{40}{E}{{}\mathbf{let}\;\Varid{q}\mathrel{=}\Varid{u}\;\mathbf{in}\;\mathbf{let}\;\Varid{p}\mathrel{=}\Varid{t}\;\mathbf{in}\;\Varid{e}{}}
\nextline[\blanklineskip]
\fromto{5}{14}{{}\mathbf{let}\;\Varid{p}\mathrel{=}{}}
\fromto{14}{35}{{}\mathbf{if}^\circ\;\Varid{t}{}}
\fromto{35}{35E}{{}\qquad\equiv\qquad{}}
\fromto{40}{E}{{}\mathbf{if}^\circ\;\Varid{t}{}}
\nextline
\fromto{14}{40}{{}\mathbf{then}\;\Varid{u}_{\mathrm{0}}\;{}}
\fromto{40}{46}{{}\mathbf{then}\;{}}
\fromto{46}{E}{{}\mathbf{let}\;\Varid{p}\mathrel{=}\Varid{u}_{\mathrm{0}}\;\mathbf{in}\;\Varid{e}{}}
\nextline
\fromto{14}{40}{{}\mathbf{else}\;\Varid{u}_{\mathrm{1}}\;{}}
\fromto{40}{46}{{}\mathbf{else}\;{}}
\fromto{46}{E}{{}\mathbf{let}\;\Varid{p}\mathrel{=}\Varid{u}_{\mathrm{1}}\;\mathbf{in}\;\Varid{e}{}}
\nextline
\fromto{5}{E}{{}\mathbf{in}\;\Varid{e}{}}
\ColumnHook
\end{pboxed}
\)\par\noindent\endgroup\resethooks
\end{itemize}
\end{definition}


We write $\G \vdash t \equiv u : \sigma$ if $\G \vdash t,u : \sigma$ and the
equation \ensuremath{\Varid{t}\equiv \Varid{u}} is derivable at the type $\sigma$.

\omitnow{
We note that the equation
\begingroup\par\noindent\advance\leftskip\mathindent\(
\begin{pboxed}\SaveRestoreHook
\column{B}{@{}l@{}}
\column{5}{@{}l@{}}
\column{54}{@{}c@{}}
\column{54E}{@{}l@{}}
\column{55}{@{}l@{}}
\column{59}{@{}l@{}}
\column{66}{@{}l@{}}
\column{E}{@{}l@{}}
\fromto{5}{54}{{}\mathbf{if}^\circ\;(\mathbf{if}^\circ\;\Varid{t}\;\mathbf{then}\;\Varid{u}_{\mathrm{0}}\;\mathbf{else}\;\Varid{u}_{\mathrm{1}})\;\mathbf{then}\;\Varid{e\char95 0}\;\mathbf{else}\;\Varid{e\char95 1}{}}
\fromto{54}{54E}{{}\qquad\equiv\qquad{}}
\fromto{59}{66}{{}\mathbf{if}^\circ\;\Varid{t}\;{}}
\fromto{66}{E}{{}\mathbf{then}\;(\mathbf{if}^\circ\;\Varid{u}_{\mathrm{0}}\;\mathbf{then}\;\Varid{e\char95 0}\;\mathbf{else}\;\Varid{e\char95 1}){}}
\nextline
\fromto{54}{55}{{}\hsindent{1}{}}
\fromto{55}{E}{{}\mathbf{else}\;(\mathbf{if}^\circ\;\Varid{u}_{\mathrm{1}}\;\mathbf{then}\;\Varid{e\char95 0}\;\mathbf{else}\;\Varid{e\char95 1}){}}
\ColumnHook
\end{pboxed}
\)\par\noindent\endgroup\resethooks
is derivable from the ones given above.

\amr{Add a lemma that if $t \perp u$, then for all environments
$\gamma$, $\evalC{t}\gamma \neq \evalC{u}\gamma$.}
}

\begin{lemma}[Soundness]
\label{lemma:soundness}
The equational theory is sound: if $\G \vdash t \equiv u : \sigma$ then the
functions $\evalC{\G \vdash t :\sigma}$ and $\evalC{\G \vdash u : \sigma}$
are extensionally equal.
\end{lemma}

\section{Completeness of the Classical Theory}
\label{sec:compclass}

The equational theory is \emph{complete} in a strong technical sense:
as we prove in the remainder of the section, any equivalence implied
by the semantics is derivable in the theory. The proof technique is
based on current work by the first author with Tarmo
Uustalu~\cite{alti:flops04}. The proof we present extends and
simplifies the method presented in that work.

\subsection{Proof Technique}

The ultimate goal is to prove the following statement.

\begin{proposition}[Completeness]
\label{prop:complete}
  If $\evalC{\G \vdash t : \sigma}$ and $\evalC{\G \vdash u : \sigma}$ are
  extensionally equal, then we can derive $\G \vdash t \equiv u : \sigma$.
\end{proposition}

In order to prove this statement, we define a function $q^\sigma_\G$ which
inverts evaluation by producing a canonical syntactical representative. In
fact, we define the function $q^\sigma_\G$ such that it maps a denotation
$\evalC{\G \vdash t : \sigma}$ to the normal form of $t$.

\begin{definition}
  The \emph{normal form} of~$t$ is given by $\nf_\G^\sigma(t) =
  q_\G^\sigma(\evalC{\G\vdash t:\sigma})$.
\end{definition}

The normal form is well-defined: given an equation $\G\vdash t \equiv
u:\sigma$, we know by soundness that $\evalC{\G\vdash t:\sigma}$ is
extensionally equal $\evalC{\G\vdash u:\sigma}$ and hence we get that
$\nf_\G^\sigma(t) = \nf_\G^\sigma(u)$. If we can now prove that the
syntactic theory can prove that every term is equal to its normal
form, then we can prove the main completeness result. Indeed given the
following lemma, we can prove completeness.

\begin{lemma}[Inversion]
\label{lemma:inversion}
The equation $\G \vdash \nf_\G^{\,\sigma}(t) \equiv t : \sigma$ is derivable.
\end{lemma}

\begin{proof*}{Proof of Proposition~\ref{prop:complete} (Completeness)}
We have:
\ba
\G \vdash t \equiv q_\G^\sigma \evalC{\G \vdash t : \sigma} : \sigma
   && \mbox{by inversion} \\
\G \vdash q_\G^\sigma \evalC{\G \vdash t : \sigma} \equiv
          q_\G^\sigma \evalC{\G \vdash u : \sigma} : \sigma
   && \mbox{by assumption} \\
\G \vdash q_\G^\sigma \evalC{\G \vdash u : \sigma} \equiv u : \sigma
   && \mbox{by inversion}
\ea
\end{proof*}

To summarise we can establish completeness by defining a function $q^\sigma_\G$
that inverts evaluation and that satisfies Inversion
Lemma~\ref{lemma:inversion}.

\subsection{Adequacy}

We begin by defining a family of functions $q^\sigma$ (``\emph{quote}'')
which invert the evaluation of \emph{closed} terms and prove a special case
of the inversion lemma for closed terms, called \emph{adequacy}. These
functions and the adequacy result are then used in the next section to invert
the evaluation of open terms and prove the general inversion lemma.



\begin{definition}
\label{def:sigma}
The \emph{syntactic representations of denotations} is given by:
\[ q^\sigma \in \evalC{\sigma} \to \ValC \sigma \]
defined by induction over $\sigma$:
\ba
q^{\Unit}\,0 &=& () \\
q^{\Qubit}\,0 &=&  \mathrm{false} \\
q^{\Qubit}\,1 &=& \mathrm{true} \\
q^{\sigma\ot\tau}\,(a,b) &=& (q^{\sigma}\,a,q^{\tau}\,b)
\ea
\end{definition}

The version of the inversion lemma for closed terms is called
\emph{adequacy}.  It guarantees that the equational theory is rich enough to
equate every closed term with its final observable value.

\begin{lemma}[Adequacy]
The equation 
  $\vdash q^{\sigma}(\evalC{~\vdash t : \sigma}~0) \equiv t :\sigma$
is derivable.
\end{lemma}
\begin{proof*}{Proof sketch.}
During the proof of such a statement we encounter open terms that must be
closed before they are ``quoted.'' So in fact the statement to prove by
induction is the following:
\[
\mbox{If~} g \in \evalC{\G} \mbox{~then~} \vdash q^{\sigma}(\evalC{\G
  \vdash t : \sigma}~g) \equiv \ensuremath{\textbf{let}^\mathbf{*}\;\Gamma\mathrel{=}q^\Gamma\;(\Varid{g})\;\mathbf{in}\;\Varid{t}} : \sigma
\]
\end{proof*}




\subsection{Inverting Evaluation}
\label{sec:c-compl}

As explained earlier, the main ingredient of the proof of completeness is the
function $q_\G^\sigma$ which inverts evaluation. To understand the basic idea
of how the inverse of evaluation is defined, consider the following example.
Let $\G$ be the environment $x : (\Qubit\ot\Qubit), y : \Qubit$ and let $f
\in \evalC{\G} \to \evalC{\Qubit}$. To find a syntactic term corresponding to
$f$, we proceed as follows:
\begin{itemize}
\item flatten all the products by introducing intermediate names; this
produces an updated environment $\G' = x_1 : \Qubit, x_2 : \Qubit, y :
\Qubit$, and an updated semantic function $f'$ such that:
\[
f'~(((((),x_1),x_2),y) = f~(((),(x_1,x_2)),y)
\]
\item enumerate all possible values for the variables, and apply $f'$ to each
enumeration to produce a result in the set $\evalC{\Qubit}$. For example, it
could be the case that $f~(((),(1,1)),1) = 0$. The result of each enumeration
can be inverted to a syntactic term using $q^\sigma$ from
Definition~\ref{def:sigma}.
\item Put things together using nested conditions representing all
the possible values for the input variables. In the example we are
considering, we get:
\begingroup\par\noindent\advance\leftskip\mathindent\(
\begin{pboxed}\SaveRestoreHook
\column{B}{@{}l@{}}
\column{5}{@{}l@{}}
\column{11}{@{}l@{}}
\column{17}{@{}l@{}}
\column{24}{@{}l@{}}
\column{30}{@{}c@{}}
\column{30E}{@{}l@{}}
\column{E}{@{}l@{}}
\fromto{B}{E}{{}\mathbf{let}\;(\Varid{x1},\Varid{x2})\mathrel{=}\Varid{x}{}}
\nextline
\fromto{B}{5}{{}\mathbf{in}\;{}}
\fromto{5}{E}{{}\mathbf{if}^\circ\;\Varid{x1}{}}
\nextline
\fromto{5}{11}{{}\mathbf{then}\;{}}
\fromto{11}{E}{{}\mathbf{if}^\circ\;\Varid{x2}{}}
\nextline
\fromto{11}{17}{{}\mathbf{then}\;{}}
\fromto{17}{24}{{}\mathbf{if}^\circ\;\Varid{y}\;{}}
\fromto{24}{30}{{}\mathbf{then}\;{}}
\fromto{30}{30E}{{}\Varid{false}{}}
\nextline
\fromto{24}{30}{{}\mathbf{else}{}}
\fromto{30}{30E}{{}\mathbin{...}{}}
\nextline
\fromto{11}{17}{{}\mathbf{else}{}}
\fromto{17}{E}{{}\mathbin{...}{}}
\nextline
\fromto{5}{11}{{}\mathbf{else}{}}
\fromto{11}{E}{{}\mathbin{...}{}}
\ColumnHook
\end{pboxed}
\)\par\noindent\endgroup\resethooks
\end{itemize}

The idea is formalised in the following definition.



\begin{definition}
\label{def:inverteval}
The function 
\[ q_\G^\sigma \in (\evalC{\G}\to \evalC{\sigma}) \to \Tmm{\G}{\sigma}\] 
for \emph{inverting evaluation} is defined by
analysing the context:
\ba
q_{\emptyC}^{\sigma}(f) &=& q^\sigma~(f~(0)) \\
q_{\G,x:\Unit}^{\sigma}(f) &=& q_{\G}^{\sigma}~(h) \qquad\mbox{where}~h(g) = f(g,0) \\
q_{\G,x:\Qubit}^{\sigma}(f) &=&
  (\rifo x \rthen q_{\G}^{\sigma}~(h_1) \relse q_{\G}^{\sigma}~(h_0)) \\
  &&\mbox{where~} h_i(g) = f(g,i) \mbox{~for~} i \in \{0,1\} \\
q_{\G,x:(\tau_1\ot\tau_2)}^{\sigma}(f) &=&
  (\rlet (x_1,x_2) = x \,\rin\, q_{\G,x_1:\tau_1,x_2:\tau_2}^{\sigma}(h) \\
  &&\mbox{where~} h(g,x_1,x_2) = f(g,(x_1,x_2))
\ea
\end{definition}
The base case is straightforward: the evaluation produces a closed
value which can be inverted using the \emph{``quote''} function of
Definition~\ref{def:sigma}. If the context includes a variable $x$ of
type $\Unit$, then we supply the only possible value for that variable
(0), and inductively construct the term with the variable $x$ bound to
$()$. The result is of the correct type because we can add or drop
bindings of variables of type $\Unit$ to the environment. If the
context includes a variable $x$ of type $\Qubit$, then we supply the
two possible values for that variable 0 and 1. A conditional is then
used to select the correct branch depending on the actual value of
$x$. Finally, if the context includes a variable of type
$\tau_1\ot\tau_2$ then we simply flatten the product and proceed
inductively. The function $q^\sigma_\G$ does indeed satisfy the
inversion lemma.


\section{Quantum Data and Control}
\label{sec:quantum}

We develop the typing rules and semantics of the quantum fragment of QML in
two stages. First we extend the judgements $\G\vdash t:\sigma$ and the
semantics of Section~\ref{sec:classsem} to handle quantum data in a
straightforward manner. This simple treatment is only however an intermediate
step in the development as it admits quantum programs that are not realisable
on a quantum computer. We then refine both the type system and the semantics
to identify exactly the realisable quantum programs. 

\subsection{The Category $\QVec$}
\label{sec:catvec}

\begin{figure*}[t]
\centering{\small
\framebox{
$\zeroR \qquad \mulR \qquad \supR$
}}
\caption{Typing quantum data (I)}
\label{fig:typingqone}
\end{figure*}

As a first approximation to a type system for QML programs, we consider the
type system of Figure~\ref{fig:typingterms} extended with the rules in
Figure~\ref{fig:typingqone}. 

Unlike the classical case, a judgement $\G\vdash t\tin\sigma$ is \emph{not}
interpreted as a function in $\evalC{\Gamma}\to\evalC{\sigma}$. Rather,
because we now have superpositions of terms with complex probability
amplitudes, we interpret such judgements as functions in
$\evalC{\Gamma}\to\evalV{\sigma}$ where $\evalV{\sigma}$ represents the
complex vectors over the base set $\evalC{\sigma}$. In other words,
$\evalV{\sigma}$ is defined to be $\evalC{\sigma}\to\Complex$ which is
sometimes denoted $\Vec{\evalC{\sigma}}$. We call the structure described
above the category $\QVec$.

Naturally this change requires that we revisit the semantics of the
classical terms given in Figure~\ref{fig:classmean} so that each
denotation returns a complex vector. For example, we should have:
\[
\evalV{\emptyC\vdash\mathrm{false}:\Qubit} = \vconst{v} 
  \quad\mbox{where~} v~0 = 1 \mbox{~and~} v~1=0
\]
Instead of mapping the value representing the empty context to the
denotation of $\mathrm{false}$, we now return a vector $v$ which
associates the denotation of $\mathrm{false}$ with probability
amplitude 1 and the denotation of $\mathrm{true}$ with probability
amplitude~0. 

This change can be done systematically by noticing that it corresponds
to a monad whose unit and lift operation are defined below:
\ba 
\vreturn{a}~(b) &=& 1 ~\mbox{if}~a=b \mbox{~and~0~otherwise} \\
f^*(v) &=& \Sigma a. (v~a) * (f~a) 
\ea
More precisely every value that is returned in Figure~\ref{fig:classmean} 
is explicitly tagged with the monadic $\vreturn{}$ and when two functions
are composed in Figure~\ref{fig:classmean} using $f \circ g$, 
the composition is replaced by $f^* \circ g$.

\begin{figure*}[t]
\centering{\small
\framebox{
$\begin{array}{rcll}
\evalV{\emptyC \vdash \overrightarrow{0} : \sigma} &=& 
  \vconst{v} & \mbox{where~} \forall a \in \evalC{\sigma}. v~a = 0 \\
\evalV{\Gamma \vdash \kappa * t : \sigma} &=& g 
  & \mbox{~where~} \begin{array}[t]{rcl}
           g~a &=& \kappa * (f a) \\
           f &=& \evalV{\G \vdash t \tin \sigma} 
               \end{array} \\
\evalV{\Gamma \vdash t+u : \sigma} &=& h 
  & \mbox{~where~} \begin{array}[t]{rcl}
               h~a &=& f~a + g~a \\
               f &=& \evalV{\G \vdash t \tin \sigma} \\
               g &=& \evalV{\G \vdash u \tin \sigma}
               \end{array} \\
\end{array}$}}
\caption{Meaning function for quantum data}
\label{fig:meaning-qone}
\end{figure*}

The meaning of the new constructs for quantum data is given in
Figure~\ref{fig:meaning-qone}.




\subsection{Orthogonality}
\label{sec:orth}
\label{sec:semantics-strict}

The type system presented so far does indeed correctly track the uses
of variables and prevents variables from being weakened; yet the
situation is more subtle. It turns out that the type system accepts
terms which implicitly perform measurements and as a consequence
accepts programs which are not realisable as quantum computations.

Consider the expression \ensuremath{\mathbf{if}^\circ\;\Varid{x}\;\mathbf{then}\;\Varid{true}\;\mathbf{else}\;\Varid{true}}: this expression
appears, syntactically at least, to use \ensuremath{\Varid{x}}. However given the
semantics of $\ifo$, which returns a superposition of the branches,
the expression happens to return \ensuremath{\Varid{true}} without really \emph{using}
any information about \ensuremath{\Varid{x}}. In order to maintain the invariant that all
measurements are explicit, the type system should reject the above
expression as well.

More precisely, the expression \ensuremath{\mathbf{if}^\circ\;\Varid{x}\;\mathbf{then}\;\Varid{t}\;\mathbf{else}\;\Varid{u}} should only be
accepted if \ensuremath{\Varid{t}} and~\ensuremath{\Varid{u}} are \emph{orthogonal} quantum values ($t\perp
u$). This notion intuitively ensures that the conditional operator
does not implicitly discard any information about \ensuremath{\Varid{x}} during the
evaluation. Because of a similar concern, the two branches of a
superposition should also be orthogonal. 

The typing rules for conditionals and superpositions are modified as
in Figure~\ref{fig:typingterms2}. This modification also achieves that
programs are normalised, \ie, the sum of the probabilities of a
superposition add up to 1.

\begin{figure*}[t]
\centering{\small
\framebox{
$\begin{array}{c}
\ifoRo \\
\supRo \\
\substR
\end{array}$}}
\caption{Typing quantum data (II)}
\label{fig:typingterms2}
\end{figure*}

In Figure~\ref{fig:orth} we define the inner product of terms, which
to any pair of terms $\Gamma\vdash t,u:\sigma$ assigns
$\ip{t}{u}\in\Complex\cup\{?\}$.  This is used to define
orthogonality: $t\perp u$ holds if $\ip{t}{u}=0$.

\begin{figure*}[t]
\centering{\small
\framebox{
$\begin{array}{rcl}
\ip{\mathrm{t}}{\mathrm{t}} &=& 1 \\
\ip{\mathrm{false}}{\mathrm{true}} &=& 0 \\
\ip{\mathrm{true}}{\mathrm{false}} &=& 0 \\
\\
\ip{\ensuremath{\overrightarrow{0}}}{\mathrm{true}} &=& 0 = \ip{\mathrm{true}}{\ensuremath{\overrightarrow{0}}}\\
\ip{\ensuremath{\overrightarrow{0}}}{\mathrm{false}}&=& 0 = \ip{\mathrm{false}}{\ensuremath{\overrightarrow{0}}}\\
\ip{\ensuremath{\overrightarrow{0}}}{\mathrm{x}}    &=& 0 = \ip{\mathrm{x}}{\ensuremath{\overrightarrow{0}}}\\
\\
\ip{(t,t')~}{~(u,u')} &=& \ip{t}{u} * \ip{t'}{u'} \\
\end{array}
\begin{array}{rcl}
\ip{\lambda*t + \lambda'*t' ~}{~u} &=&
  \lambda^* * \ip{t}{u} + \lambda'^* * \ip{t'}{u} \\
\ip{t~}{~\kappa*u + \kappa'*u' } &=&
  \kappa * \ip{t}{u} + \kappa' * \ip{t}{u'} \\
\\
\ip{\lambda*t}{u} &=& \conj{\lambda}\ip{t}{u}\\
\ip{t}{\lambda*u} &=& \lambda\ip{t}{u}\\
\ip{t+t'}{u} &=& \ip{t}{u}+\ip{t'}{u}\\
\ip{t}{u+u'} &=& \ip{t}{u}+\ip{t}{u'}\\
\\
\ip{t}{u} &=& ? \qquad\qquad\mbox{otherwise}
\end{array}$}}
\caption{Inner products and orthogonality}
\label{fig:orth}
\end{figure*}

The judgement $\vdasho$ is not automatically closed under the equality
judgement, hence we add the rule (subst). Our philosophy is that we
allow equivalent representations of QML programs which do not satisfy
the orthogonality criteria locally, as long as the program as a whole
is equivalent to one which does satisfy the criteria.

\subsection{The Category $\Qo$}
\label{sec:catq0}

The restriction of the set of typable terms requires a similar
semantic restriction. All we need to do is to restrict the morphisms
in the category of complex vectors to satisfy the following two
conditions:
\begin{itemize}
\item Linearity: If $f \in \Vec{A}\to\Vec{B}$, $\alpha\in\Complex$,
and $v, v_1, v_2 \in \Vec{A}$, then $f (v_1 + v_2) = f(v_1) + f(v_2)$
and $f (\alpha v) = \alpha (f~v)$.
\item Isometry: If $f \in \Vec{A}\to\Vec{B}$ and $v_1, v_2 \in
\Vec{A}$, then $\ip{v_1}{v_2} = \ip{f~v_1}{f~v_2}$. (In other words,
$f$ preserves inner products of vectors.)
\end{itemize}
Two morphisms $f,g \in A \to B$ are \emph{orthogonal} if for all vector $v
\in \Vec{A}$, we have $\ip{f~v}{g~v} = 0$.  We call the resulting category,
the category $\Qo$ of strict quantum computations.  The homset of morphisms
in $\evalC{\Gamma}\to\evalV{\sigma}$ satisfying the above conditions is
called $\Qom{\evalC{\G}}{\evalQ{\sigma}}$.


\omitnow{
\jjg{Code added for T. below}
\begingroup\par\noindent\advance\leftskip\mathindent\(
\begin{pboxed}\SaveRestoreHook
\column{B}{@{}l@{}}
\column{3}{@{}l@{}}
\column{E}{@{}l@{}}
\fromto{3}{E}{{}\Varid{v}\mathbin{+}\Varid{w}\bind \Varid{f}\mathrel{=}\Varid{v}\bind \Varid{f}\mathbin{+}\Varid{w}\bind \Varid{f}{}}
\nextline
\fromto{3}{E}{{}\Varid{v}\bind \Varid{f}\mathbin{+}\Varid{g}\mathrel{=}\Varid{v}\bind \Varid{f}\mathbin{+}\Varid{v}\bind \Varid{y}{}}
\nextline
\fromto{3}{E}{{}\Varid{v}\mathbin{*}\Varid{f}\bind \Varid{h}\mathrel{=}\Varid{v}\bind \Varid{f}\mathbin{*}\Varid{h}{}}
\ColumnHook
\end{pboxed}
\)\par\noindent\endgroup\resethooks
}


The meaning function is given as before but with the maps interpreted
in the category $\Qo$, \ie, the meaning of a derivation $\Gamma\vdash
t \tin \sigma$ is a morphism $\evalQ{\Gamma\vdash t \tin
\sigma}\in\Qom{\evalC{\G}}{\evalQ{\sigma}}$. The requirement for
orthogonality in the type system is reflected semantically: for
isometries $f,g$, we have that $\cond{f}{g}$ is an isometry, if $f$
and $g$ are orthogonal.

\omitnow{
\begin{figure*}[t]
\centering{\small
\framebox{
$\begin{array}{rcl}
\evalQ{{\emptyC \vdash () \tin \Unit}} &=& \vreturn{0} \\
\evalQ{{\emptyC \vdash \false \tin \Qubit}} &=& \vreturn{0} \\
\evalQ{{\emptyC \vdash \true \tin \Qubit}} &=& \vreturn{1} \\
\evalQ{x \tin \sigma \vdash x \tin \sigma} &=& \mathit{id} \\
\evalQ{\G\ot\D \vdash \rlet x = t \,\rin\, u \tin \tau} &=&
  g \circ (f \ot \mathit{id}) \circ \delta_{\G,\D} \\
  && \mbox{~where~}
  \begin{array}[t]{rcl}
   f &=& \evalQ{\G \vdash t \tin \sigma} \\
   g &=& \evalQ{\D,x\tin\sigma \vdash t \tin \tau}
               \end{array} \\
\evalQ{\G \ot \D \vdash (t,u) \tin \sigma \ot \tau} &=&
  (f \ot g) \circ \delta_{\G,\D} \\
  && \mbox{~where~} \begin{array}[t]{rcl}
               f &=& \evalQ{\G \vdash t \tin \sigma} \\
               g &=& \evalQ{\D \vdash u \tin \tau}
               \end{array} \\
\\
\evalQ{\G\ot\D \vdash \rlet (x,y) = t \rin u \tin \rho} &=&
  g \circ (f \ot \mathit{id}) \circ \delta_{\G,\D} \\
  && \mbox{~where~}
  \begin{array}[t]{rcl}
   f &=& \evalQ{\G \vdash t \tin \sigma \ot \tau} \\
   g &=& \evalQ{\D,\, x \tin \sigma, y \tin \tau \vdash u \tin \rho}
               \end{array} \\
\\
\evalQ{\G\ot\D \vdash \rifo c \rthen t \relse u \tin \sigma} &=&
  (\condo{f}{g}) \circ (h \ot \mathit{id}) \circ \delta_{\G,\D} \\
  && \mbox{~where~}
  \begin{array}[t]{rcl}
    h &=& \evalQ{\G \vdash c \tin \Qubit} \\
    f &=& \evalQ{\D \vdash t \tin \sigma} \\
    g &=& \evalQ{\D \vdash u \tin \sigma}
  \end{array}
\\
\evalQ{\emptyC \vdash \overrightarrow{0} : \sigma} &=& \overrightarrow{0} \\
\\
\evalQ{\Gamma \vdash \kappa * t : \sigma} &=& \kappa * f \\
  && \mbox{~where~} f = \evalQ{\G \vdash t \tin \sigma} \\
\\
\evalQ{\Gamma \vdash t+u : \sigma} &=& \ot~(f+g) \\
  && \mbox{~where~} \begin{array}[t]{rcl}
               f &=& \evalQ{\G \vdash t \tin \sigma} \\
               g &=& \evalQ{\G \vdash u \tin \sigma}
               \end{array} \\
\end{array}$}}
\caption{Meaning function for strict computations}
\label{fig:meaning-strict}
\end{figure*}
}

\subsection{Quantum Equational Theory}
\label{sec:quantumcomp}

The equational theory for the quantum language inherits all the
equations for the classical case. This can be informally verified by
noting that the meaning function in the case of the quantum language
is essentially identical to the classical case. Formally, the proof
technique explained in Section~\ref{sec:classcomplete} applies equally
well to the quantum case and yields the same equations for the
classical core plus additional equations to deal with quantum data.


\begin{definition}
The \emph{quantum equations} are:
  \begin{description}
  \item[(\ensuremath{\mathbf{if}^\circ})] \mbox{}
    \begingroup\par\noindent\advance\leftskip\mathindent\(
\begin{pboxed}\SaveRestoreHook
\column{B}{@{}c@{}}
\column{BE}{@{}l@{}}
\column{6}{@{}l@{}}
\column{E}{@{}l@{}}
\fromto{6}{E}{{}\mathbf{if}^\circ\;(\lambda\mathbin{*}\Varid{t}_{\mathrm{0}}\mathbin{+}\kappa\mathbin{*}\Varid{t}_{\mathrm{1}})\;\mathbf{then}\;\Varid{u}_{\mathrm{0}}\;\mathbf{else}\;\Varid{u}_{\mathrm{1}}{}}
\nextline
\fromto{B}{BE}{{}\qquad\equiv\qquad{}}
\fromto{6}{E}{{}\lambda\mathbin{*}(\mathbf{if}^\circ\;\Varid{t}_{\mathrm{0}}\;\mathbf{then}\;\Varid{u}_{\mathrm{0}}\;\mathbf{else}\;\Varid{u}_{\mathrm{1}})\mathbin{+}\kappa\mathbin{*}(\mathbf{if}^\circ\;\Varid{t}_{\mathrm{1}}\;\mathbf{then}\;\Varid{u}_{\mathrm{0}}\;\mathbf{else}\;\Varid{u}_{\mathrm{1}}){}}
\ColumnHook
\end{pboxed}
\)\par\noindent\endgroup\resethooks
  \item[(superpositions)]\mbox{}
    \begingroup\par\noindent\advance\leftskip\mathindent\(
\begin{pboxed}\SaveRestoreHook
\column{B}{@{}l@{}}
\column{8}{@{}l@{}}
\column{26}{@{}l@{}}
\column{E}{@{}l@{}}
\fromto{8}{26}{{}\Varid{t}\mathbin{+}\Varid{u}{}}
\fromto{26}{E}{{}\qquad\equiv\qquad\Varid{u}\mathbin{+}\Varid{t}{}}
\nextline
\fromto{8}{26}{{}\Varid{t}\mathbin{+}\overrightarrow{0}{}}
\fromto{26}{E}{{}\qquad\equiv\qquad\Varid{t}{}}
\nextline
\fromto{8}{26}{{}\Varid{t}\mathbin{+}(\Varid{u}\mathbin{+}\Varid{v}){}}
\fromto{26}{E}{{}\qquad\equiv\qquad(\Varid{t}\mathbin{+}\Varid{u})\mathbin{+}\Varid{v}{}}
\nextline
\fromto{8}{26}{{}\lambda\mathbin{*}(\Varid{t}\mathbin{+}\Varid{u}){}}
\fromto{26}{E}{{}\qquad\equiv\qquad\lambda\mathbin{*}\Varid{t}\mathbin{+}\lambda\mathbin{*}\Varid{u}{}}
\nextline
\fromto{8}{26}{{}\lambda\mathbin{*}\Varid{t}\mathbin{+}\kappa\mathbin{*}\Varid{t}{}}
\fromto{26}{E}{{}\qquad\equiv\qquad(\lambda\mathbin{+}\kappa)\mathbin{*}\Varid{t}{}}
\nextline
\fromto{8}{26}{{}\mathrm{0}\mathbin{*}\Varid{t}{}}
\fromto{26}{E}{{}\qquad\equiv\qquad\overrightarrow{0}{}}
\ColumnHook
\end{pboxed}
\)\par\noindent\endgroup\resethooks
  \end{description}
\end{definition}

\begin{lemma}[Soundness]
\label{lemma:soundness2}
The equational theory is sound: if $\G \vdash t \equiv u : \sigma$ then the
isometries $\evalQ{\G \vdash t :\sigma}$ and $\evalQ{\G \vdash u : \sigma}$
are extensionally equal.
\end{lemma}

The additional equations are used to prove equality between different
quantum values. Semantically, two quantum values are the same if they
denote the same vector, which is the case if the sum of the paths to
each classical value is the same. For example, to find a simplified
quantum value equivalent to:
\begingroup\par\noindent\advance\leftskip\mathindent\(
\begin{pboxed}\SaveRestoreHook
\column{B}{@{}l@{}}
\column{E}{@{}l@{}}
\fromto{B}{E}{{}(\Varid{false}\mathbin{+}\Varid{true})\mathbin{+}(\Varid{false}\mathbin{+}(\mathbin{-}\mathrm{1})\mathbin{*}\Varid{true}){}}
\ColumnHook
\end{pboxed}
\)\par\noindent\endgroup\resethooks
we first normalise to:
\begingroup\par\noindent\advance\leftskip\mathindent\(
\begin{pboxed}\SaveRestoreHook
\column{B}{@{}l@{}}
\column{3}{@{}l@{}}
\column{E}{@{}l@{}}
\fromto{3}{E}{{}(\mathrm{1}\mathbin{/}\sqrt{\mathrm{2}})\mathbin{*}((\mathrm{1}\mathbin{/}\sqrt{\mathrm{2}})\mathbin{*}\Varid{false}\mathbin{+}(\mathrm{1}\mathbin{/}\sqrt{\mathrm{2}})\mathbin{*}\Varid{true})\mathbin{+}{}}
\nextline
\fromto{3}{E}{{}(\mathrm{1}\mathbin{/}\sqrt{\mathrm{2}})\mathbin{*}((\mathrm{1}\mathbin{/}\sqrt{\mathrm{2}})\mathbin{*}\Varid{false}\mathbin{+}(\mathbin{-}\mathrm{1}\mathbin{/}\sqrt{\mathrm{2}})\mathbin{*}\Varid{true}){}}
\ColumnHook
\end{pboxed}
\)\par\noindent\endgroup\resethooks
This term has two paths to \ensuremath{\Varid{false}}; along each of them the product of
the amplitudes is \ensuremath{(\mathrm{1}\mathbin{/}\sqrt{\mathrm{2}})\mathbin{*}(\mathrm{1}\mathbin{/}\sqrt{\mathrm{2}})} which is \ensuremath{\mathrm{1}\mathbin{/}\mathrm{2}}. The sum of
all the paths to \ensuremath{\Varid{false}} is \ensuremath{\mathrm{1}}, and the sum of all the paths to \ensuremath{\Varid{true}}
is \ensuremath{\mathrm{0}}. In other words, the entire term is equivalent to simply
\ensuremath{\Varid{false}}. The above calculation proves that the Hadamard operation is
self-inverse, as discussed in the introduction.

\subsection{Quoting quantum values}
\label{sec:qcomplete}

We will now adapt the techniques developed in section
\ref{sec:classcomplete} to the quantum case. A classical value
$v\in\ValC\sigma$ is simply a term representing an element in
$\evalC{\sigma}$. A quantum value represents a vector in
$\Vec{\evalQ{\sigma}}$, hence we have to close values under
superpositions. We define $\ValQ\,\sigma\subseteq \CTm\,\sigma$ inductively as a subset of
closed terms of type $\sigma$:
\begin{itemize}
\item $\ru{v\in\ValC\,\sigma}{\val\,v\in\ValQ\,\sigma}$

\item $0 \in \ValQ\,\sigma$
\item $\ru{v,w\in\ValQ\,\sigma}{v + w\in\ValQ\,\sigma}$

\item $\ru{v\in\ValQ\,\sigma}{\kappa*v \in\ValQ\,\sigma}$
\end{itemize}
We write $\ValQo\sigma$ for isometric quantum values which satisfy 
the restrictions introduced in Figure \ref{fig:typingterms2}.

We have already seen that there is a monadic structure on $\Vec{A}=A
\to\Complex$. Correspondingly, we have a Kleisli structure on $\ValQ$;
$\val\in\ValC\sigma \to \ValQ\sigma$ is the return and bind is defined
as given $v\in\ValQ\sigma$ and $f\in\ValC\sigma \to
\ValQ\tau$, we define $\ensuremath{\Varid{v}\bind \Varid{f}} \in \ValQ\,\tau$ by induction over
$v$:
\begingroup\par\noindent\advance\leftskip\mathindent\(
\begin{pboxed}\SaveRestoreHook
\column{B}{@{}l@{}}
\column{3}{@{}l@{}}
\column{25}{@{}l@{}}
\column{E}{@{}l@{}}
\fromto{3}{25}{{}(\Varid{val}\;\Varid{x}){}}
\fromto{25}{E}{{}\bind \Varid{f}\mathrel{=}\Varid{f}\;\Varid{x}{}}
\nextline
\fromto{3}{25}{{}\mathrm{0}{}}
\fromto{25}{E}{{}\bind \Varid{f}\mathrel{=}\mathrm{0}{}}
\nextline
\fromto{3}{25}{{}\Varid{v}\mathbin{+}\Varid{w}{}}
\fromto{25}{E}{{}\bind \Varid{f}\mathrel{=}(\Varid{v}\bind \Varid{f})\mathbin{+}(\Varid{w}\bind \Varid{f}){}}
\nextline
\fromto{3}{25}{{}\kappa\mathbin{*}\Varid{v}{}}
\fromto{25}{E}{{}\bind \Varid{f}\mathrel{=}\kappa\mathbin{*}(\Varid{v}\bind \Varid{f}){}}
\ColumnHook
\end{pboxed}
\)\par\noindent\endgroup\resethooks
\begin{lemma}
  $(\ValC,\ValQ,val,\ensuremath{(\bind )})$ is a Kleisli structure, i.e. it satisfies
  the following equations:
  \begin{enumerate}
  \item \ensuremath{\Varid{val}\;\Varid{x}\bind \Varid{f}\equiv \Varid{f}\;\Varid{x}}

  \item \ensuremath{\Varid{v}\bind \lambda \Varid{x}.\Varid{val}\;\Varid{x}\equiv \Varid{v}}

  \item \ensuremath{\Varid{v}\bind \lambda \Varid{x}.(\Varid{f}\;\Varid{x})\bind \Varid{g}\equiv (\Varid{v}\bind \Varid{f})\bind \Varid{g}}
  \end{enumerate}
\end{lemma}
\begin{proof}
  Case (i) follows from the definition. Cases (ii) and (iii) can be
  shown by induction over the structure of $v$.
\end{proof}

While the classical definition of $q^\sigma$ (def. \ref{def:sigma}) was
completely straightforward, its quantum counterpart is a bit more subtle, in
particular the in the case of tensor products. As a special case consider
$q^{\Qubit\ot\Qubit}$, given an element
\[\vecx{v}\in\evalQ{\Qubit\ot\Qubit} = \evalC{\Qubit}\times\evalC{\Qubit} \to \Complex\]
we have to construct a value $q^{\Qubit\ot\Qubit}\,\vecx{v}\in\ValQ\,\Qubit\ot\Qubit$.
This can be done by calculating the probabilities that the first qubit
is $i$, $\fst\,\vecx{v}\,i\in\RealPlus$, given by
\[\fst\,\vecx{v}\,i = \sqrt{|\vecx{v}(i,0)|^2+|\vecx{v}(i,1)|^2}\]
creating the first level of
the value as a tree, and then for the second level normalising the amplitudes
wrt. the probabilities of the previous level, see figure \ref{fig:valtree}
for the corresponding tree.
\begin{figure*}[t]
\centering{\large
\framebox{
\xymatrix@C=2em{
&&&\ar[dll]_{\ensuremath{\Varid{fst}\;\Varid{v}\;\mathrm{0}}}\ar[drr]^{\ensuremath{\Varid{fst}\;\Varid{v}\;\mathrm{1}}}&&&\\
&\ar[dl]_{\frac{v(0,0)}{\ensuremath{\Varid{fst}\;\Varid{v}\;\mathrm{0}}}}\ar[dr]^{\frac{v(0,1)}{\ensuremath{\Varid{fst}\;\Varid{v}\;\mathrm{0}}}}&&&
&\ar[dl]_{\frac{v(1,0)}{\ensuremath{\Varid{fst}\;\Varid{v}\;\mathrm{1}}}}\ar[dr]^{\frac{v(1,1)}{\ensuremath{\Varid{fst}\;\Varid{v}\;\mathrm{1}}}}\\
\mbox{\small (0,0)}&&\mbox{\small (0,1)}&&\mbox{\small (1,0)}&&\mbox{\small (1,1)}\\
}}}
\caption{Value tree for $\Qubit\ot\Qubit$}
\label{fig:valtree}
\end{figure*}
We write $\evalP{\sigma} = \evalC{\sigma}\to\RealPlus$ for the set of probability 
distributions, obviously we have $\evalP{\sigma}\subseteq\evalQ{\sigma}$. We 
observe that  $\fst\,\vecx{v}\in \evalP{\sigma}$.
Generalising the idea given above we arrive at the following definition
of quote:
\begin{definition}
\label{def:qqsigma}
The \emph{syntactic representations of denotations} is given by
\[ q^\sigma \in \evalQ{\sigma} \to \ValQ\,\sigma \]
defined by induction over $\sigma$:
\ba
q^{\Unit}\,\vecx{v} &=& (\vecx{v}\,0) * ()
\\ q^{\Qubit}\,\vecx{v} &=& (\vecx{v}\,1) * \true + (\vecx{v}\,0)* \false
\\ q^{\sigma\ot\tau}\,\vecx{v} &=&
q^\sigma(\fst\,\vecx{v}) \\
&& \qquad \ensuremath{\bind } \lambda x\in\evalC{\sigma}.(\pinv{(\fst\,\vecx{v})}\,x)*q^\tau(\lambda y.\vecx{v}(x,y))\\
&& \qquad \ensuremath{\bind \lambda \Varid{y}.\Varid{val}\;(\Varid{x},\Varid{y})}
\ea
where:
\ba
\fst & \in & \evalQ{\sigma\ot\tau} \to \evalP{\sigma} \\
\fst\, \vecx{v}\,x & = & \sqrt{\Sigma y.|\vecx{v}(x,y)|^2}\\
\\
\pinv{-} & \in & \evalP{\sigma} \to {\evalP{\sigma}}\\
\pinv{\vecx{v}}\,x & = & \ensuremath{\lambda \Varid{x}.\mathbf{if}\;\Varid{p}\;\Varid{x}\equiv \mathrm{0}\;\mathbf{then}\;\mathrm{0}\;\mathbf{else}\;\mathrm{1}\mathbin{/}(\Varid{p}\;\Varid{x})}
\ea
\end{definition}

To show adequacy we have to establish a number of properties of $q^\sigma$: we have
to show that it is linear and isometric and that it preserves tensor products. This is 
summarised in the following proposition:

\begin{proposition}\label{prop:q-lin}\mbox{}
  \begin{enumerate}
  \item $q^\sigma\,(\kappa * \vecx{v}) \ensuremath{\equiv } \kappa * (q^\sigma\,\vecx{v})$

  \item $q^\sigma\,(\vecx{v} + \vecx{w}) \ensuremath{\equiv } (q^\sigma\,\vecx{v}) + (q^\sigma\,\vecx{w})$

  \item $\ip{\vecx{v}}{\vecx{w}} = \ip{q^\sigma\,\vecx{v}}{q^\sigma\,\vecx{w}}$

  \item $q^{\sigma\ot\tau}\,(\vecx{v}\ot \vecx{w}) \equiv (q^\sigma\,\vecx{v},q^\tau\,\vecx{w})$
  \end{enumerate}

\end{proposition}

The proof of the above proposition again isn't completely straightforward, e.g. linearity
cannot just be proven by induction over $\sigma$. It is essential that we first 
establish some properties of renormalising a vector wrt. a probability distribution.
We define the product of a probability distribution $p\in\evalP{\sigma}$
and a vector $\vecx{v}\in\evalQ{\sigma}$ as:
\ba
p*\vecx{v} & \in & \evalQ{\sigma}\\
p*\vecx{v} & = & \lambda x\in\evalC{\sigma}.(p x)*(\vecx{v}\, x)
\ea
It is not hard to see that an analogous operation can be defined
on values, given $v\in\ValQ\,\sigma$ and $p\in\evalP{\sigma}$ as above,
we define:
\ba
p*v & \in & \ValQ\,\sigma\\
p*v & = & v \ensuremath{\bind } \lambda x\in\evalC{\sigma}.(p x)*(\val\, x)
\ea
The key property we establish is
\begin{lemma}\label{lem:heureka}
  Given $p\in\evalP{\sigma}$ and $\vecx{v}\in\evalQ{\sigma}$
  \[ p*(q^\sigma\,\vecx{v}) \ensuremath{\equiv } q^\sigma\,(p*\vecx{v}) \]
\end{lemma}
which can be verified by induction over $\sigma$ and observing that 
while $\pinv{-}$ isn't a proper inverse, it nevertheless satisfies the 
following property
\[ \pinv{(p+q)}*(p+q) = (\pinv{p})*p \]

Using the fact that $q^\sigma$ is isometric we can show that it produces
values satisfying the orthogonality constraints:
\begin{proposition}
  Given $v\in \evalQ{\sigma}$
  \[ \vdasho q^\sigma\,v :\sigma\]
\end{proposition}

\omitnow{
\begin{lemma}\label{lem:bindplus}
\ensuremath{\Varid{v}\bind \Varid{f}\mathbin{+}\Varid{g}\equiv (\Varid{v}\bind \Varid{f})\mathbin{+}(\Varid{v}\bind \Varid{g})}
\end{lemma}
\begin{proof}
By induction over \ensuremath{\Varid{v}}.
\end{proof}
%

We will show that as in the classical case, $\evalQ{-}$ has an inverse
$q^\sigma$. As a tool we need to normalise vectors by probability
distributions. We define the space of probability distributions
$\evalP{\sigma} = \evalC{\sigma} \to \RealPlus$ and we view this as
a subspace of the vector space $\evalP{\sigma}\subseteq\evalC{\sigma}$.
We define the product of a probability distribution $p\in\evalP{\sigma}$
and a vector $\vecx{v}\in\evalC{\sigma}$ as:
\ba
p*\vecx{v} & \in & \evalQ{\sigma}\\
p*\vecx{v} & = & \lambda x\in\evalC{\sigma}.(p x)*(\vecx{v}\, x)
\ea
It is not hard to see that an analogous operation can be defined
on values, given $v\in\ValQ\,\sigma$ and $p\in\evalP{\sigma}$ as above,
we define:
\ba
p*v & \in & \ValQ\,\sigma\\
p*v & = & v \ensuremath{\bind } \lambda x\in\evalC{\sigma}.(p x)*(\val\, x)
\ea
We note that this multiplication commutes with the Kleisli-bind
operator both for vectors, $\vecx{v}\in\evalC{\sigma}$, and values;
$v$ and $p$ as above.

\begin{lemma}\label{lem:prob-bind}
~
\[
\begin{array}{l}
  p * \vecx{v} \ensuremath{\bind \Varid{f}\mathrel{=}}\vecx{v}\ensuremath{\bind }\lambda x.(p x)*(f x)\\
  p * v \ensuremath{\bind \Varid{f}\mathrel{=}\Varid{v}\bind } \lambda x.(p x)*(f x)
\end{array}
\]
\end{lemma}
It will also be useful to have a partial inverse distribution, that is
given $p\in\evalP{\sigma}$ we define:
\ba
\pinv{p} & \in & \evalP{\sigma}\\
\pinv{p} & = & \ensuremath{\lambda \Varid{x}.\mathbf{if}\;\Varid{p}\;\Varid{x}\equiv \mathrm{0}\;\mathbf{then}\;\mathrm{0}\;\mathbf{else}\;\mathrm{1}\mathbin{/}\Varid{p}\;\Varid{x}}
\ea
We have that $\pinv{p}$ is only a partial inverse, \ie, it is not the case that
$(\pinv{p})*p$ is the constant $1$ distribution, because this depends on
where $p$ is $0$. However, the following fact will turn out to be useful.
\begin{lemma}\label{lem:pinv-lem}
  Let $p,q\in\evalP{\sigma}$:
  \[ \pinv{(p+q)}*(p+q) = (\pinv{p})*p \]
\end{lemma}

To lift $q$ over tensor products we define an operation $\fst$
which calculates a probability distribution depending on the first
projection. Given $v\in\evalQ{\sigma\ot\tau}$ we define:
\ba
\fst\, v & \in & \evalP{\sigma} \\
\fst\, v & = & \lambda x.\sqrt{\Sigma y.|v(x,y)|^2}
\ea
We are now ready to define  $q^\sigma\in\evalQ{\sigma}\to\ValQ\,\sigma$ by
induction over $\sigma$:
\ba q^{\Unit}\,v &=& (v\,0) * ()
\\ q^{\Qubit}\,v &=& (v\,1) * \true + (v\,0)* \false
\\ q^{\sigma\ot\tau}\,v &=&
q^\sigma(\fst\,v) \ensuremath{\bind } \lambda x\in\evalC{\sigma}.((\pinv{\fst\,v})\,x)*q^\tau(\lambda y.v(x,y))\ensuremath{\bind \lambda \Varid{y}.\Varid{val}\;(\Varid{x},\Varid{y})}
\ea

To establish that $q$ is linear and isometric, it is necessary to show
that it commutes with $p*-$.

\begin{lemma}\label{lem:heureka}
  \[ p*(q^\sigma\,\vecx{v}) \ensuremath{\equiv } q^\sigma\,(p*\vecx{v}) \]
\end{lemma}





\begin{proposition}\label{prop:q-lin}
  $q^\sigma$ is linear and isometric and preserves $\ot$, that is:
  \begin{enumerate}
  \item $q^\sigma\,(\kappa * \vecx{v}) \ensuremath{\equiv } \kappa * (q^\sigma\,\vecx{v})$

  \item $q^\sigma\,(v + w) \ensuremath{\equiv } (q^\sigma\,v) + (q^\sigma\,w)$

  \item $\ip{v}{w} = \ip{q^\sigma\,v}{q^\sigma\,w}$

  \item $q^{\sigma\ot\tau}\,v\ot w \equiv (q^\sigma\,v,q^\tau\,w)$
  \end{enumerate}

\end{proposition}
\begin{proof} For ii) we reason as follows, by induction over $\sigma$:
\[
\begin{array}{l}
q^{\sigma \ot \tau} (v+w)\\
= \{ \mbox{definition of $q$}\}\\
q^{\sigma} (\fst\, (v+w)) \ensuremath{\bind \lambda \Varid{x}.} (\pinv{\fst\, (v+w)\, x})*q^{\tau}(\ensuremath{\lambda \Varid{y}.\Varid{v}\mathbin{+}\Varid{w}\;(\Varid{x},\Varid{y}))\bind \lambda \Varid{y}.\Varid{val}\;(\Varid{x},\Varid{y})}\\
\ensuremath{\equiv } \{ \mbox{induction hypothesis} \}\\
q^{\sigma} (\fst\, (v+w)) \ensuremath{\bind } \\
  \ensuremath{\lambda \Varid{x}.}(\pinv{\fst\, (v+w)\, x})*(q^{\tau}(\ensuremath{\lambda \Varid{y}.\Varid{v}\;(\Varid{x},\Varid{y}))\mathbin{+}}q^{\tau}(\ensuremath{\lambda \Varid{y}.\Varid{w}\;(\Varid{x},\Varid{y})))\bind \lambda \Varid{y}.\Varid{val}\;(\Varid{x},\Varid{y})}\\
\ensuremath{\equiv } \{ \mbox{lemma \ref{lem:bindplus}} \}\\
q^{\sigma} (\fst\, (v+w)) \ensuremath{\bind \lambda \Varid{x}.}(\pinv{\fst\, (v+w)\, x})*(q^{\tau}(\ensuremath{\lambda \Varid{y}.\Varid{v}\;(\Varid{x},\Varid{y})))\bind \lambda \Varid{y}.\Varid{val}\;(\Varid{x},\Varid{y})}\\
  + q^{\sigma} (\fst\, (v+w)) \ensuremath{\bind \lambda \Varid{x}.}(\pinv{\fst\, (v+w)\, x})*(q^{\tau}(\ensuremath{\lambda \Varid{y}.\Varid{w}\;(\Varid{x},\Varid{y})))\bind \lambda \Varid{y}.\Varid{val}\;(\Varid{x},\Varid{y})}\\
\ensuremath{\equiv } \{ \mbox{lemmas \ref{lem:prob-bind}, \ref{lem:pinv-lem}, \ref{lem:heureka}} \}\\
q^{\sigma} (\fst\, v) \ensuremath{\bind \lambda \Varid{x}.}(\pinv{\fst\, v\, x})*(q^{\tau}(\ensuremath{\lambda \Varid{y}.\Varid{v}\;(\Varid{x},\Varid{y})))\bind \lambda \Varid{y}.\Varid{val}\;(\Varid{x},\Varid{y})}\\
  + q^{\sigma} (\fst\, w) \ensuremath{\bind \lambda \Varid{x}.}(\pinv{\fst\, w\, x})*(q^{\tau}(\ensuremath{\lambda \Varid{y}.\Varid{w}\;(\Varid{x},\Varid{y})))\bind \lambda \Varid{y}.\Varid{val}\;(\Varid{x},\Varid{y})}\\
= \{\mbox{definition of $q$}\}\\
(q^{\sigma \ot \tau} v)+(q^{\sigma \ot \tau} w)
\end{array}
\]
The proof for i) follows similar lines.
\end{proof}



}

\subsection{Adequacy}

We define a syntactic counterpart to:
\[\delta_{\G,\D}\in\Qom{\evalC{\Gamma\ot\Delta}}{(\evalQ{\Gamma}\ot\evalQ{\Delta})}\]
as:
\[\deltax_{\G,\D}\in\Tmm{(\Gamma\ot\Delta)}{(\con{\Gamma}\ot\con{\Delta})}\]
by:
\[
\deltax_{\G,\D} =
 \left\{ \begin{array}{rl}
  \ensuremath{\mathbf{let}\;(\Varid{g},\Varid{d})\mathrel{=}} \delta_{\G',\D'} \ensuremath{\mathbf{in}\;((\Varid{g},\Varid{x}),(\Varid{d},\Varid{x}))} &
    \mbox{if~} \G=\G',x:\sigma \\
    \qquad    \mbox{~and~}       \D=\D',x:\sigma \\
  \ensuremath{\mathbf{let}\;(\Varid{g},\Varid{d})\mathrel{=}} \delta_{\G',\D} \ensuremath{\mathbf{in}\;((\Varid{g},\Varid{x}),\Varid{d})} &
    \mbox{if~} \G=\G',x:\sigma \\
    \qquad \mbox{~and~} x \not\in \dom{\,\Delta} \\
  1_\D & \mbox{if~} \G=\emptyC
   \end{array}\right.
\]

To establish that $q^\sigma$ commutes with the context operations we
have to show that contraction corresponds to
$\delta\in\Qom{\evalC{\sigma}}{(\evalQ{\sigma}\ot\evalQ{\sigma})}$.

\begin{lemma}
  Given $v\in \evalQ{\sigma}$ we have
  \[\ensuremath{\mathbf{let}\;\Varid{x}\mathrel{=}}q^\sigma\,v ~\ensuremath{\mathbf{in}\;(\Varid{x},\Varid{x})\equiv }q^{\sigma\ot\sigma}\,v \]
\end{lemma}
\begin{proof}
  By induction on $\sigma$.
\end{proof}

Exploiting this property we can show that the context operations commute with
quote:
\begin{lemma}\label{lem:delta-lem}
  Given $\vecx{v}\in\evalQ{\G\ot\D}$
  \[q^{\con{\G}\ot\con{\D}}\,(\delta_{\G,\D}\,{\vecx{v}}) \equiv
  \deltax_{\G,\D}\,q^{\con{\G\ot\D}}\,\vecx{v}\]
\end{lemma}

\begin{theorem}
\label{theorem:qadequacy}
If $\G \vdash t : \sigma$ and $g \in \evalQ{\G}$ then
\[\vdash q^{\sigma}(\evalQ{\G \vdash t : \sigma}g) \equiv
\ensuremath{\textbf{let}^\mathbf{*}\;\Gamma\mathrel{=}q^\Gamma\;\Varid{g}\;\mathbf{in}\;\Varid{t}} : \sigma.\]
\end{theorem}
\begin{proof}
By induction over the derivation of $\G \vdash t : \sigma$, as an example
consider the case for let:
\[\begin{array}{l}
q^\rho\,(\evalQ{\G\ot\D \vdash \rlet x = t \rin u \tin \rho})\\
\equiv \{ \mbox{definition of $\evalQ{\dots}$} \}\\
q^\rho\,(\evalQ{u} \circ (\evalQ{t} \ot \mathit{id}) \circ \delta_{\G,\D})\\
\equiv \{ \mbox{induction hypothesis for $u$ and $t$} \}\\
u\circ (t\circ q^\G \ot q^\D ) \circ \delta_{\G,\D})\\
\equiv  \{ \mbox { lemma \ref{lem:delta-lem} } \} \}\\
u\circ (t\ot \mathit{id}) \circ \deltax_{\G,\D}\circ q^{\con{\G\ot\D}}\\
\equiv \\
(\rlet x = t \rin u)\circ q^{\con{\G\ot\D}}
\end{array}\]
The other cases use the same style of reasoning to deal with the structural
properties and exploit proposition \ref{prop:q-lin}. Note that the case for
\ensuremath{\mathbf{if}^\circ} can be reduced to linearity.
\end{proof}

\begin{corollary}[Adequacy]\label{cor:q-adeq}
  If $\vdash t : \sigma$ then
  $\vdash q^{\sigma}(\evalQ{~\vdash t : \sigma}) \equiv t :\sigma$
\end{corollary}

\subsection{Completeness and normalisation}

The development here follows closely the one in the classical case as
presented in Section \ref{sec:c-compl}.

\begin{definition}
\label{def:qinverteval}
The function:
\[q_\G^\sigma\in\Qom{\evalC{\G}}{\evalQ{\sigma}}\to\Tmm{\G}{\sigma}\]
for \emph{inverting evaluation} is defined by 
analysing the context:
\ba
q_{\emptyC}^{\sigma}(f) &=& q^\sigma~(f~(\vreturn 0)) \\
q_{\G,x:\Unit}^{\sigma}(f) &=& 
\phi^{-1}_{\G,x:\Unit} \circ
  (q_{\G}^\rho) \circ
  \Phi_{\G,x:\Unit} \\
q_{\G,x:\Qubit}^{\sigma}(f) &=&
  \phi^{-1}_{\G,x:\Qubit} \circ
  (q_\G^\sigma \times q_\G^\sigma) \circ
  \Phi_{\G,x:\Qubit} \\
q_{\G,x:(\tau_1\ot\tau_2)}^{\sigma}(f) &=&
  \phi^{-1}_{\G,x:\tau_1\ot\tau_2} \circ
  q_{\G,x_1:\tau_1,x_2:\tau_2}^\sigma \circ
  \Phi_{\G,x:\tau_1\ot\tau_2}
\ea

The auxiliary isomorphisms are defined as follows:
\ba
\phi_{\G,x:\Unit} & \in & \Tmm{(\G,x:\Unit)}{\sigma}\to\Tmm{\G}{\sigma}\\
\phi_{\G,x:\Unit} t & = & \rlet x = () \,\rin\, t \\
\phi_{\G} t & = & t\\
\\
\phi_{\G,x:\Qubit} & \in & \Tmm{(\G,x:\Qubit}{\sigma}) \to \{(t_0,t_1)\in(\Tmm{\G}{\sigma})^2\mid t_0\perp t_1\} \\
\phi_{x:\Qubit}\,t &=& (\rlet x = \mathrm{false} \,\rin\, t,
                         \rlet x = \mathrm{true} \,\rin\, t) \\
\phi^{-1}_{\G,x:\Qubit} (t,u) &=& \rifo x \rthen t \relse u \\
\\
\phi_ {\G,x:\tau_1\ot\tau_2}& \in & \Tmm{(\G,x:\tau_1\ot\tau_2)}{\rho}\to\Tmm{(\G,x_1:\tau_1,x_2:\tau_2)}\\
\phi_{\G,x:\tau_1\ot\tau_2}\,t &=& \rlet x = (x_1,x_2) \,\rin\, t \\
\phi^{-1}_{\G,x:\tau_1\ot\tau_2} (t) &=& \rlet (x_1,x_2) = x \,\rin\, t
\ea
 The semantic map corresponding to each $\phi$ is written $\Phi$.
\end{definition}

For the inversion proof we only need the provability of one side of the 
isomorphisms which follows from the $\eta$-equalities.

\begin{lemma}\label{lem:sem-eta}
  The following family of equalities is derivable
  \[ \phi^{-1}_\G (\phi_\G t) \equiv t \]
\end{lemma}

\begin{definition}
  The \emph{normal form} of~$t$ is given by $\nf_\G^\sigma(t) =
  q_\G^\sigma(\evalQ{\G\vdash t:\sigma})$.
\end{definition}

\begin{lemma}[Inversion]
\label{lemma:qinversion}
The equation $\G \vdash \nf_\G^{\,\sigma}(t) \equiv t$ is derivable.
\end{lemma}
\begin{proof}
  By induction over the definition of $q_\G^\sigma$. In the case of $\G=\emptyC$ the
  result follows from adequacy, Corollary \ref{cor:q-adeq}. In all the other cases
  we exploit Lemma \ref{lem:sem-eta}.
\end{proof}

Since all our definitions are effective $\nf$ indeed gives rise to a normalisation 
algorithm. As a consequence, our equational theory is decidable, modulo deciding
equalities of the complex number terms which occur in our programs. We also note that 
as in the classical case, our theory is complete:

\begin{proposition}[Completeness]
\label{prop:qcomplete}
  If $\evalQ{\G \vdash t : \sigma}$ and $\evalQ{\G \vdash u : \sigma}$ are
  extensionally equal, then we can derive $\G \vdash t \equiv u : \sigma$.
\end{proposition}

\section{Conclusions and Further Work}
\label{sec:conc}

We have developed a sound and complete equational theory for a
functional quantum programming language, while at the same time
providing a normalisation algorithm. The construction is a modular
extension of a classical theory, indeed the quantum theory inherits
not just all the equations and term formers, it is also possible
to generalise our proof technique to the quantum case. The quantum 
theory introduces additional constructs corresponding to superpositions and 
equations relating them. 

The obvious next step is to generalise this approach to the full language QML
including measurements. The equational theory is already a challenge, since 
a measurement can have non-local effects on shared data. Semantically, we
will be using superoperators to model programs with measurements. Clearly, 
we have to extend our quote operator to work on density matrices.

Another interesting direction, would be to consider higher order quantum
programs and develop a complete equational theory and normalisation algorithm
for this calculus. A likely semantic domain is given by presheaves, here the
tensor product can be modelled using Day's construction, which is
automatically closed, \ie, provides an interpretation for higher types.

\omitnow{
\appendix

\section{Proof of Lemma \ref{lemma:soundness}}
\label{app:soundness}

\begin{itemize}
\item $\beta$-equations
\begin{enumerate}




\item \begingroup\par\noindent\advance\leftskip\mathindent\(
\begin{pboxed}\SaveRestoreHook
\column{B}{@{}l@{}}
\column{26}{@{}c@{}}
\column{26E}{@{}l@{}}
\column{30}{@{}l@{}}
\column{E}{@{}l@{}}
\fromto{B}{26}{{}\mathbf{let}\;(\Varid{x},\Varid{y})\mathrel{=}(\Varid{t},\Varid{u})\;\mathbf{in}\;\Varid{e}{}}
\fromto{26}{26E}{{}\equiv {}}
\fromto{30}{E}{{}\mathbf{let}\;\Varid{x}\mathrel{=}\Varid{t}\;\mathbf{in}\;\mathbf{let}\;\Varid{y}\mathrel{=}\Varid{u}\;\mathbf{in}\;\Varid{e}.{}}
\ColumnHook
\end{pboxed}
\)\par\noindent\endgroup\resethooks

First, the left hand side.

\[
\begin{array}{rl}
\mbox{~lhs~} &= \evalx{{\G\ot\D \ot \D' \vdash \rlet (x,y) = (t,u) \rin e \tin \rho}} \\
& = g_e \circ (f_{(t,u)} \times \mathit{id}) \circ \delta_{(\G \ot \D),\D'}
\end{array}
\]

\noindent where

\[
\begin{array}{rcl}
   g_e &=& \evalx{{\D',\, x \tin \sigma, y \tin \tau \vdash u \tin \rho}}\\
   f_{(t,u)} &=& \evalx{{\G \ot \D \vdash (t,u) \tin \sigma \ot \tau}}\\
   &=& (f_t \circ f_u) \circ \delta_{\G, \D}\\
   \mbox{~where~} &&\\
   f_t &=& \evalx{{\G \vdash t \tin \sigma}}\\
   f_u &=& \evalx{{\D \vdash u \tin \tau}}
\end{array}
\]

Then, we have:

\[
\begin{array}{rl}
\mbox{~lhs~} = &  g_e \circ (((f_t \circ f_u)) \circ \delta_{\G,\D}) \times \mathit{id}) \circ \delta_{(\G \ot \D),\D'}
\end{array}
\]

To the right hand side:

\[
\begin{array}{rl}
\mbox{~rhs~} &= \evalx{{\G \ot\D \ot \D' \vdash \rlet x = t \rin \rlet y = u  \rin e \tin \rho}} \\
&=g_{let} \circ (f_t \times \mathit{id}) \circ \delta_{\G, (\D \ot \D')}
\end{array}
\]

\noindent where

\[
\begin{array}{rcl}
   f_t &=& \evalx{{\G \vdash t \tin \sigma}}\\
   g_{let} &=& \evalx{{\D \ot \D', x\tin \sigma \vdash \rlet y = u \rin e \tin \rho }}\\
   &=& g_e \circ (f_u \times \mathit{id}) \circ \delta_{\D,\D'}\\
   \mbox{~where~} &&\\
   f_u &=& \evalx{{ \D \vdash u \tin \tau }}\\
   g_e &=& \evalx{{\D', x \tin \sigma, y \tin \tau \vdash e \tin \rho }}
\end{array}
\]

Then, we have:

\[
\begin{array}{rl}
\mbox{~rhs~}&=
g_e \circ (f_u \times \mathit{id}) \circ \delta_{\D,\D'} \circ (f_t \times \mathit{id})
\circ \delta_{\G,(\D \ot \D')}
\end{array}
\]

Finally, because parallel composition is extensionally equal to sequential
composition using identity in the extra wires, we have that $\mbox{~lhs~}$ is
extensionally equal to $\mbox{~rhs~}$.

\item \begingroup\par\noindent\advance\leftskip\mathindent\(
\begin{pboxed}\SaveRestoreHook
\column{B}{@{}l@{}}
\column{27}{@{}c@{}}
\column{27E}{@{}l@{}}
\column{31}{@{}l@{}}
\column{E}{@{}l@{}}
\fromto{B}{27}{{}\mathbf{if}^\circ\;\Varid{false}\;\mathbf{then}\;\Varid{t}\;\mathbf{else}\;\Varid{u}{}}
\fromto{27}{27E}{{}\equiv {}}
\fromto{31}{E}{{}\Varid{u}{}}
\ColumnHook
\end{pboxed}
\)\par\noindent\endgroup\resethooks

First the left hand side:

\[
\begin{array}{rl}
 \mbox{~lhs~} &= \evalx{{\G\ot\D \vdash \rifo false \rthen t \relse u \tin \sigma}} \\
&=    (\cond{g}{h}) \circ (f \times \mathit{id}) \circ \delta_{\G,\D}
\end{array}
\]

\noindent where

\[
\begin{array}{rcl}
    f &=& \evalx{{\emptyC\vdash false \tin \Qubit}} \\
     &=&  \vreturn{2}\\
    g &=& \evalx{{\D \vdash t \tin \sigma}} \\
    h &=& \evalx{{\D \vdash u \tin \sigma}}
\end{array}
\]

Then, using the conditional definition and because
$\G$ is empty:

\[
\begin{array}{r}
\mbox{~lhs~}= (\cond{g}{h}) \circ (\vreturn{2}
\times \mathit{id}) \circ \mathit{id} = h
\end{array}
\]

Now, the right hand side:

\[
\begin{array}{l}
\mbox{~rhs~} = \evalx{{ \emptyC \ot \D  \vdash u \tin \sigma}} = h.
\end{array}
\]

\item \begingroup\par\noindent\advance\leftskip\mathindent\(
\begin{pboxed}\SaveRestoreHook
\column{B}{@{}l@{}}
\column{26}{@{}c@{}}
\column{26E}{@{}l@{}}
\column{30}{@{}l@{}}
\column{E}{@{}l@{}}
\fromto{B}{26}{{}\mathbf{if}^\circ\;\Varid{true}\;\mathbf{then}\;\Varid{t}\;\mathbf{else}\;\Varid{u}{}}
\fromto{26}{26E}{{}\equiv {}}
\fromto{30}{E}{{}\Varid{u}{}}
\ColumnHook
\end{pboxed}
\)\par\noindent\endgroup\resethooks

First the left hand side:

\[
\begin{array}{rl}
 \mbox{~lhs~} &= \evalx{{\G\ot\D \vdash \rifo true \rthen t \relse u \tin \sigma}} \\
&=    (\cond{g}{h}) \circ (f \times \mathit{id}) \circ \delta_{\G,\D}
\end{array}
\]

\noindent where

\[
\begin{array}{rcl}
    f &=& \evalx{{\emptyC\vdash true \tin \Qubit}} \\
     &=&  \vreturn{1}\\
    g &=& \evalx{{\D \vdash t \tin \sigma}} \\
    h &=& \evalx{{\D \vdash u \tin \sigma}}
\end{array}
\]

Then, using the conditional definition and because
$\G$ is empty:

\[
\begin{array}{r}
\mbox{~lhs~}= (\cond{g}{h}) \circ (\vreturn{1}
\times \mathit{id}) \circ \mathit{id} = g
\end{array}
\]

Now, the right hand side:

\[
\begin{array}{l}
\mbox{~rhs~} = \evalx{{ \emptyC \ot \D  \vdash t \tin \sigma}} = g.
\end{array}
\]
\end{enumerate}

\item $\eta$-equations
\begin{enumerate}
\item \begingroup\par\noindent\advance\leftskip\mathindent\(
\begin{pboxed}\SaveRestoreHook
\column{B}{@{}l@{}}
\column{29}{@{}c@{}}
\column{29E}{@{}l@{}}
\column{33}{@{}l@{}}
\column{E}{@{}l@{}}
\fromto{B}{29}{{}\mathbf{let}\;\Varid{x}\mathrel{=}\Varid{t}\;\mathbf{in}\;\Varid{x}{}}
\fromto{29}{29E}{{}\equiv {}}
\fromto{33}{E}{{}\Varid{t}{}}
\ColumnHook
\end{pboxed}
\)\par\noindent\endgroup\resethooks

\[
\begin{array}{rl}
 \mbox{~lhs~} &= \evalx{{\G\ot\D \vdash \rlet x = t \rin x \tin \sigma}} \\
&=  g \circ   (f \times \mathit{id}) \circ \delta_{\G,\D}
\end{array}
\]

\noindent where

\[
\begin{array}[t]{rcl}
    f &=& \evalx{{\G \vdash t \tin \sigma}} \\
    g &=& \evalx{{\D, x \tin \sigma \vdash  x \tin \sigma}} \\
    &=& \evalx{{\emptyC, x \tin \sigma \vdash  x \tin \sigma}} \\
    &=& \mathit{id}
\end{array}
\]

\noindent then

\[
\begin{array}{rl}
\mbox{~lhs~} & = \mathit{id}  \circ   (f \times \mathit{id}) \circ \delta_{\emptyC,\D}\\
&= f
\end{array}
\]

Using the fact that $\D$ is empty, we have:

\[
\begin{array}{rl}
\mbox{~rhs~} &=  \evalx{{\G \ot \emptyC \vdash t \tin \sigma}}\\
&=  \evalx{{\G \vdash t \tin \sigma}}\\
&= f
\end{array}
\]

\item \begingroup\par\noindent\advance\leftskip\mathindent\(
\begin{pboxed}\SaveRestoreHook
\column{B}{@{}l@{}}
\column{28}{@{}c@{}}
\column{28E}{@{}l@{}}
\column{32}{@{}l@{}}
\column{E}{@{}l@{}}
\fromto{B}{28}{{}\mathbf{let}\;(\Varid{x},\Varid{y})\mathrel{=}\Varid{t}\;\mathbf{in}\;(\Varid{x},\Varid{y}){}}
\fromto{28}{28E}{{}\equiv {}}
\fromto{32}{E}{{}\Varid{t}{}}
\ColumnHook
\end{pboxed}
\)\par\noindent\endgroup\resethooks

\[
\begin{array}{rl}
 \mbox{~lhs~} &= \evalx{{\G\ot\D \vdash \rlet (x,y) = t \rin (x,y) \tin \rho}} \\
&=  g \circ   (f \times \mathit{id}) \circ \delta_{\G,\D}
\end{array}
\]

\noindent where

\[
\begin{array}[t]{rcl}
    f &=& \evalx{{\G \vdash t \tin \sigma \ot \tau}} \\
    g &=& \evalx{{\D, x \tin \sigma, y \tin \tau  \vdash  (x,y) \tin \rho}} \\
    &=& \evalx{{\emptyC, x \tin \sigma, y \tin \tau  \vdash  (x,y) \tin \rho}} \\
    &=& (g_x \times g_y) \circ \delta_{x,y}\\
    \mbox{~where~}&&\\
    g_x &=& \evalx{{ x \tin \sigma \vdash x \tin \sigma}} = \mathit{id}\\
    g_y &= & \evalx{{ y \tin \tau \vdash y \tin \tau}} = \mathit{id}
\end{array}
\]

\noindent then

\[
\begin{array}{rcl}
\mbox{~lhs~} &=& (\mathit{id} \times \mathit{id}) \circ \delta_{x,y}
\circ ( f \times \mathit{id}) \circ \delta_{\G,\emptyC}\\
&=& f
\end{array}
\]

Using the facts that $\D$ is empty and that $\rho \equiv \sigma
\ot \tau$, we have:

\[
\begin{array}{rl}
\mbox{~rhs~} &=  \evalx{{\G \ot \emptyC \vdash t \tin \rho}}\\
&=  \evalx{{\G \vdash t \tin \rho}}\\
&= f
\end{array}
\]

\item \begingroup\par\noindent\advance\leftskip\mathindent\(
\begin{pboxed}\SaveRestoreHook
\column{B}{@{}l@{}}
\column{E}{@{}l@{}}
\fromto{B}{E}{{}\mathbf{if}^\circ\;\Varid{t}\;\mathbf{then}\;\Varid{true}\;\mathbf{else}\;\Varid{false}\equiv \Varid{t}{}}
\ColumnHook
\end{pboxed}
\)\par\noindent\endgroup\resethooks

\[
\begin{array}{rcl}
\mbox{~lhs}&=& \evalx{{ \G \ot \D \vdash \rifo t \rthen \mathrm{true} \relse
\mathrm{false} \tin \Qubit }}\\
&=&(\cond{g}{h}) \circ (f \times \mathit{id}) \circ \delta_{\G,\D}
\end{array}
\]

\noindent where

\[
\begin{array}{rcl}
    f &=& \evalx{{\G \vdash t \tin \Qubit}} \\
    g &=& \evalx{{\emptyC \vdash \mathrm{true} \tin \Qubit}} \\
     &=& \vreturn{1}\\
    h &=& \evalx{{\emptyC \vdash \mathrm{false} \tin \Qubit}}\\
     &=& \vreturn{2}\\
\end{array}
\]

Then, using the conditional definition and because
$\D$ is empty:

\[
\begin{array}{rcl}
\mbox{~lhs~}&=& (\cond{\vreturn{1}}{\vreturn{2}})
\circ (f \times \mathit{id}) \circ \delta_{\G,\emptyC}\\
&=& f\\
&=& \evalx{{ \G \ot \emptyC \vdash t \tin \Qubit}}\\
&=& \mbox{~rhs~}
\end{array}
\]

\end{enumerate}

\item Commuting conversions

\begin{enumerate}
\item \begingroup\par\noindent\advance\leftskip\mathindent\(
\begin{pboxed}\SaveRestoreHook
\column{B}{@{}l@{}}
\column{5}{@{}l@{}}
\column{12}{@{}l@{}}
\column{E}{@{}l@{}}
\fromto{B}{E}{{}\mathbf{let}\;\Varid{x}\mathrel{=}\mathbf{if}^\circ\;\Varid{t}\;\mathbf{then}\;\Varid{u0}\;\mathbf{else}\;\Varid{u1}\;\mathbf{in}\;\Varid{e}{}}
\nextline
\fromto{B}{5}{{}\equiv {}}
\fromto{5}{12}{{}\mathbf{if}^\circ\;\Varid{t}\;{}}
\fromto{12}{E}{{}\mathbf{then}\;\mathbf{let}\;\Varid{x}\mathrel{=}\Varid{u0}\;\mathbf{in}\;\Varid{e}{}}
\nextline
\fromto{12}{E}{{}\mathbf{else}\;\mathbf{let}\;\Varid{x}\mathrel{=}\Varid{u1}\;\mathbf{in}\;\Varid{e}{}}
\ColumnHook
\end{pboxed}
\)\par\noindent\endgroup\resethooks

\[
\begin{array}{rcl}
\mbox{~lhs~} &=& \llbracket \G \ot \D \ot \D' \vdash \rlet x = \rifo t \\
&& \;\;\rthen u_o \relse u_1 \rin e \tin \tau \rrbracket \\
&=& g_e \circ (f_{if} \times \mathit{id}) \circ \delta_{(\G \ot \D),\D'}
\end{array}
\]

\noindent where

\[
\begin{array}{rcl}
f_{if} &=& \evalx{{\G \ot \D \vdash \rifo t \rthen u_o \relse u_1 \tin \sigma}}\\
&=&  (\cond{f_{u0}}{f_{u1}}) \circ (f_t \times \mathit{id}) \circ \delta_{\G,\D}\\
f_{t} &=& \evalx{{ \G \vdash t \tin \Qubit}}\\
f_{u0} &=& \evalx{{ \D \vdash u_0 \tin \sigma}}\\
f_{u1} &=& \evalx{{ \D \vdash u_1 \tin \sigma}}\\
g_e &=& \evalx{{\D', x \tin \sigma \vdash e \tin \tau }}
\end{array}
\]

\noindent Therefore:

\[
\begin{array}{rcl}
\mbox{~lhs~} &=& g_e \circ (((\cond{f_{u0}}{f_{u1}}) \circ (f_t \times \mathit{id}) \circ
\delta_{\G,\D}) \times \mathit{id}) \circ \\
&&\delta_{(\G \ot \D), \D'}
\end{array}
\]

Now, the right hand side:

\[
\begin{array}{rcl}
\mbox{~rhs~} &=& \llbracket \G \ot \D \ot \D' \vdash \rifo t \\
&&\;\; \rthen \rlet x = u_0 \rin e \\
&& \;\;\relse \rlet x = u_1\rin e \tin \tau \rrbracket \\
&=& (\cond{g_{let1}}{h_{let2}}) \circ (f_t \times \mathit{id}) \circ
\delta_{\G,(\D \ot \D')}
\end{array}
\]

\noindent where

\[
\begin{array}{rcl}
f_t &=& \evalx{{\G \vdash t \tin \Qubit }}\\
g_{let1} &=& \evalx{{\D \ot \D' \vdash \rlet x = u_0 \rin e \tin \tau }} \\
&=& g_e \circ (f_{u0} \times \mathit{id}) \circ \delta_{\D,\D'}\\
f_{u0}&=&\evalx{{\D \vdash u_0 \tin \sigma }}\\
g_e &=& \evalx{{\D' , x \tin \sigma \vdash e \tin \tau  }}\\
h_{let2} &=& \evalx{{\D \ot \D' \vdash \rlet x = u_1 \rin e \tin \tau }} \\
&=&  g_e \circ (f_{u1} \times \mathit{id}) \circ \delta_{\D,\D'}\\
f_{u1}&=&\evalx{{\D \vdash u_1 \tin \sigma }}\\
\end{array}
\]

\noindent Therefore:

\[
\begin{array}{rcl}
\mbox{~rhs~} &=& ((g_e \circ (f_{u0} \times \mathit{id}) \circ \delta_{\D,\D'}) |\\
&&(g_e \circ (f_{u1} \times \mathit{id}) \circ \delta_{\D,\D'})) \\
&&\circ (f_t \times \mathit{id})  \circ \delta_{\G,(\D \ot \D')}
\end{array}
\]

\noindent
which is extensionally equal to:

\[
\begin{array}{rcl}
\mbox{~lhs~} &=& g_e \circ (((\cond{f_{u0}}{f_{u1}}) \circ (f_t \times \mathit{id}) \circ
\delta_{\G,\D}) \times \mathit{id}) \circ \\
&&\delta_{(\G \ot \D), \D'}
\end{array}
\]
\item \begingroup\par\noindent\advance\leftskip\mathindent\(
\begin{pboxed}\SaveRestoreHook
\column{B}{@{}c@{}}
\column{BE}{@{}l@{}}
\column{5}{@{}l@{}}
\column{E}{@{}l@{}}
\fromto{5}{E}{{}\mathbf{let}\;\Varid{x}\mathrel{=}\Varid{t}\;\mathbf{in}\;(\Varid{u0},\Varid{u1}){}}
\nextline
\fromto{B}{BE}{{}\equiv {}}
\fromto{5}{E}{{}(\mathbf{let}\;\Varid{x}\mathrel{=}\Varid{t}\;\mathbf{in}\;\Varid{u0},\mathbf{let}\;\Varid{x}\mathrel{=}\Varid{t}\;\mathbf{in}\;\Varid{u1}){}}
\ColumnHook
\end{pboxed}
\)\par\noindent\endgroup\resethooks

\[
\begin{array}{rcl}
\mbox{~lhs} &=& \evalx{{\G \ot \D \ot \D' \vdash \rlet x = t \rin (u_o,u_1) \tin \sigma \ot \tau}}\\
&=& g_{(u_0,u_1)} \circ (f_t \times \mathit{id}) \circ \delta_{\G,(\D \ot \D')}
\end{array}
\]

\noindent where

\[
\begin{array}{rcl}
f_t &=& \evalx{{\G \vdash t \tin \rho }}\\
g_{(u_o,u_1)}&=& \evalx{{ (\D\ot\D'), x\tin \rho \vdash (u_0,u_1) \tin \sigma \ot \tau}}\\
&=& (g_{u0} \times g_{u1}) \circ \delta_{(\D,x \tin \rho),(\D',x \tin \rho)}\\
g_{uo} &=& \evalx{{\D , x \tin \rho \vdash u_0 \tin \sigma}}\\
g_{u1}&=& \evalx{{\D' , x \tin \rho \vdash u_1 \tin \tau }}
\end{array}
\]

Therefore:

\[
\begin{array}{rcl}
\mbox{~lhs~}&=& (g_{u0} \times g_{u1}) \circ \delta_{(\D,x \tin \rho), (\D',x \tin \rho)}
\circ (f_t \times \mathit{id}) \circ \\
&&\delta_{\G,(\D \ot\D')}
\end{array}
\]

Now the right hand side:

\[
\begin{array}{rcl}
\mbox{~rhs~}&=& \llbracket \G \ot \D \ot \D' \vdash \\
&&(\rlet x = t \rin u_0,
\rlet x = t \rin u_1) \tin \sigma \ot \tau \rrbracket\\
&&(f_{let1} \times g_{let2}) \circ \delta_{(\G \ot \D),(\G \ot \D')}
\end{array}
\]

\noindent where

\[
\begin{array}{rcl}
f_{let1} &=& \evalx{{\G \ot \D \vdash \rlet x = t \rin u_0 \tin \sigma }}\\
&=& g_{u0} \circ (f_t \times \mathit{id} ) \circ \delta_{\G,\D}\\
g_{u0} &=& \evalx{{ \D , x \tin \rho \vdash u_0 \tin \sigma}}\\
f_t &=& \evalx{{\G \vdash t \tin \rho }}\\
g_{let2} &=& \evalx{{ G \ot \D' \vdash \rlet x = t \rin u_1 \tin \tau}}\\
&=&  g_{u1} \circ (f_t \times \mathit{id} ) \circ \delta_{\G,\D'}\\
g_{u1} &=& \evalx{{ \D' , x \tin \rho \vdash u_1 \tin \tau}}\\
\end{array}
\]

Then

\[
\begin{array}{rcl}
\mbox{~rhs~}&=& ((g_{u0} \circ (f_t \times \mathit{id}) \circ \delta_{\G,\D}) \times  \\
&&(g_{u1} \circ (f_t \times \mathit{id}) \circ \delta_{\G,\D'}))\circ
\delta_{(\G \ot \D),(\G \ot \D')}
\end{array}
\]

\item \begingroup\par\noindent\advance\leftskip\mathindent\(
\begin{pboxed}\SaveRestoreHook
\column{B}{@{}c@{}}
\column{BE}{@{}l@{}}
\column{5}{@{}l@{}}
\column{12}{@{}l@{}}
\column{E}{@{}l@{}}
\fromto{5}{E}{{}\mathbf{if}^\circ\;(\mathbf{if}^\circ\;\Varid{t}\;\mathbf{then}\;\Varid{u0}\;\mathbf{else}\;\Varid{u1})\;\mathbf{then}\;\Varid{e0}\;\mathbf{else}\;\Varid{e1}{}}
\nextline
\fromto{B}{BE}{{}\equiv {}}
\fromto{5}{12}{{}\mathbf{if}^\circ\;\Varid{t}\;{}}
\fromto{12}{E}{{}\mathbf{then}\;(\mathbf{if}^\circ\;\Varid{u0}\;\mathbf{then}\;\Varid{e0}\;\mathbf{else}\;\Varid{e1}){}}
\nextline
\fromto{12}{E}{{}\mathbf{else}\;(\mathbf{if}^\circ\;\Varid{u1}\;\mathbf{then}\;\Varid{e0}\;\mathbf{else}\;\Varid{e1}){}}
\ColumnHook
\end{pboxed}
\)\par\noindent\endgroup\resethooks

\[
\begin{array}{rcl}
\mbox{~lhs~} &=& \llbracket \G \ot \D \ot D'\vdash \rifo \\
&&(\rifo t \rthen u_0 \relse u_1) \\
&&\rthen e_0 \relse e_1 \tin \sigma \rrbracket\\
&=& (\cond{g_{e0}}{g_{e1}}) \circ (f_{if} \times \mathit{id} ) \circ \delta_{\G\ot\D,\D'}
\end{array}
\]

\noindent where

\[
\begin{array}{rcl}
f_{if}&=& \evalx{{ \G \ot \D \vdash \rifo t \rthen u_0 \relse u_1 \tin \Qubit}}\\
&=&\cond{(f_{u0}}{f_{u1}} \circ (f_t \times \mathit{id}) \circ \delta_{\G,\D}\\
f_t &=& \evalx{{\G \vdash t \tin \Qubit }}\\
f_{u0} &=& \evalx{{\D \vdash u_0 \tin \Qubit }}\\
f_{u1} &=& \evalx{{\D \vdash u_1 \tin \Qubit }}\\
g_{eo}&=& \evalx{{\D'\vdash e_0 \tin \sigma }}\\
g_{e1}&=& \evalx{{\D'\vdash e_1 \tin \sigma}}
\end{array}
\]

Then

\[
\begin{array}{rcl}
\mbox{~lhs~} &=& (\cond{g_{e0}}{g_{e1}}) \circ (((\cond{f_{uo}}{f_{u1}})
\circ (f_t \times \mathit{id}) \circ \delta_{\G,\D}) \\
&&\times \mathit{id} ) \circ \delta_{\G\ot\D,\D'}.
\end{array}
\]

Now, the right hand side:

\[
\begin{array}{rcl}
\mbox{~rhs~} &=& \llbracket \G \ot \D \ot \D'\vdash \rifo t \rthen \\
&&(\rifo u_0 \rthen e_0 \relse e_1)\rrbracket \\
&& (\rifo u_1 \rthen e_0 \relse e_1)\\
&=& (\cond{g}{h}) \circ (f_t \times \mathit{id}) \circ \delta_{\G,(\D\ot\D')}
\end{array}
\]

\noindent where

\[
\begin{array}{rcl}
g&=& \evalx{{ \D \ot \D'\vdash \rifo u_0 \rthen e_0 \relse e_1 \tin \sigma}}\\
&=& (\cond{g_{e0}}{g_{e1}}) \circ (f_{u0} \times \mathit{id}) \circ \delta_{\D,\D'}\\
h&=& \evalx{{ \D \ot \D'\vdash \rifo u_1 \rthen e_0 \relse e_1 \tin \sigma}}\\
&=& (\cond{g_{e0}}{g_{e1}}) \circ (f_{u1} \times \mathit{id}) \circ \delta_{\D,\D'}\\
\end{array}
\]

\[
\begin{array}{rcl}
\mbox{~rhs~} &=& (\cond{((\cond{g_{e0}}{g_{e1}}) \circ (f_{u0} \times \mathit{id}) \circ
\delta_{\D,\D'})} \\
&&{((\cond{g_{e0}}{g_{e1}})\circ (f_{u0} \times \mathit{id}) \circ \delta_{\D,\D'})}) \\
&&\circ (f_t \times \mathit{id}) \circ  \delta_{\G,(\D\ot\D')}.
\end{array}
\]

\end{enumerate}
\end{itemize}

\section{Proof of Lemma \ref{lemma:adequacy}}
\label{app:adeq}

The proof is by induction on the derivation:

\begin{enumerate}
\item If $\emptyC \vdash () \tin \Unit$ and $g \in
\evalx{{ \emptyC}} = \evalx{{ \Unit}} = \{ 0\}$ then we want to show
that $q^{\Unit} (\evalx{{ \emptyC \vdash () \tin \Unit}} g)
\equiv close ((),\emptyC, q^{\emptyC}(g)) \tin \Unit$.

\[
\begin{array}{rcl}
q^{\Unit} (\evalx{{ \emptyC \vdash \mathrm{false} \tin \Unit}} g)&\equiv&
q^{\Unit} (\mathit{id} \;0)\\
&\equiv& q^{\Unit} (0) \\
&\equiv& ()\\
&\equiv& close ((),\emptyC, ())\\
&\equiv& close ((),\emptyC, q^{\emptyC}(g)).
\end{array}
\]

\item If $\emptyC \vdash \mathrm{false} \tin \Qubit$ and $g \in
\evalx{{ \emptyC}} = \evalx{{ \Unit}} = \{ 0\}$ then we want to show
that $q^{\Qubit} (\evalx{{ \emptyC \vdash \mathrm{false} \tin \Qubit}} g)
\equiv close (\mathrm{false},\emptyC, q^{\emptyC}(g)) \tin \Qubit$.

\[
\begin{array}{rcl}
q^{\Qubit} (\evalx{{ \emptyC \vdash \mathrm{false} \tin \Qubit}} g)&\equiv&
q^{\Qubit} (\vreturn{2} \;0)\\
&\equiv& q^{\Qubit} (2) \\
&\equiv& \mathrm{false}\\
&\equiv& close (\mathrm{false},\emptyC, ())\\
&\equiv& close (\mathrm{false},\emptyC, q^{\emptyC}(g)).
\end{array}
\]

\item If $\emptyC \vdash \mathrm{true} \tin \Qubit$ and $g \in
\evalx{{ \emptyC}} = \evalx{{ \Unit}} = \{ 0\}$ then we want to show
that $q^{\Qubit} (\evalx{{ \emptyC \vdash \mathrm{true} \tin \Qubit}} g)
\equiv close (\mathrm{true},\emptyC, q^{\emptyC}(g)) \tin \Qubit$.

\[
\begin{array}{rcl}
q^{\Qubit} (\evalx{{ \emptyC \vdash \mathrm{true} \tin \Qubit}} g)&\equiv&
q^{\Qubit} (\vreturn{1} \;0)\\
&\equiv& q^{\Qubit} (1) \\
&\equiv& \mathrm{true}\\
&\equiv& close (\mathrm{true},\emptyC, ())\\
&\equiv& close (\mathrm{true},\emptyC, q^{\emptyC}(g)).
\end{array}
\]

\item If $x \tin \sigma \vdash x \tin \sigma$ and $g \in
\evalx{{ x \tin \sigma}} = \evalx{{ \sigma}}$. Then we want to show
that $q^{\sigma} (\evalx{{x \tin \sigma \vdash x \tin \sigma }} \; s)
= close(x, x\tin \sigma,q^{x \tin \sigma}(s))$.

\[
\begin{array}{rcl}
q^{\sigma} (\evalx{{x \tin \sigma \vdash x \tin \sigma }} \; s) &=&
q^{\sigma} (\mathit{id} \; s) =q^{\sigma} (s)
\end{array}
\]

\noindent and

\[
\begin{array}{rcl}
close(x, x\tin \sigma,q^{\sigma}(s))&=&
\rlet x = q^{\sigma} (s) \rin close(x, \emptyC,())\\
&=& \rlet x = q^{\sigma} (s) \rin x\\
&=&  q^{\sigma} (s)
\end{array}
\]

\item If $\G \ot \D \vdash (t,u) \tin \sigma \ot \tau$ and $g \in \evalx{{ \G \ot
\D}}$ then we want to show that $q^{\sigma \ot \tau}
(\evalx{{\G \ot \D \vdash (t,u) \tin \sigma \ot \tau}} \; g) \equiv
close((t,u), \G \ot \D, q^{\G \ot \D}(g))\tin \sigma \ot \tau$. We will prove
by induction over context definition.

\begin{enumerate}
\item $\G \ot \D = \emptyC$. Then $g \in \{0\}$.

\[
\begin{array}{l}
q^{\sigma \ot \tau}
= (\evalx{{\emptyC \vdash (t,u) \tin \sigma \ot \tau}} \; g)\\
= q^{\sigma \ot \tau} ((f \times g) \circ \delta_{\emptyC,\emptyC}) \;0\\
= q^{\sigma \ot \tau} (f \times g) (0,0) \\
= q^{\sigma \ot \tau} (f(0), g(0)) \\
= (q^{\sigma} f(0), q^{\tau} g(0))
\end{array}
\]

\noindent where

\[
\begin{array}{rcl}
f &=& \evalx{{ \emptyC \vdash t \tin \sigma}}\\
g&=& \evalx{{ \emptyC \vdash u \tin \tau}}
\end{array}
\]

Using induction (on the main statement of Lemma 2).

\[
\begin{array}{rcl}
q^{\sigma} (f(0)) &=& close(t,\emptyC,())\tin \sigma\\
&=& t\\
q^{\tau} (g(0)) &=& close(u,\emptyC,()) \tin \tau\\
&=&u
\end{array}
\]

Then

\[
\begin{array}{rcl}
(q^{\sigma} f(0), q^{\tau} g(0))&=& (t,u)\\
&=& close((t,u), \emptyC, ()) \\
&=& close((t,u), \emptyC, q^{\emptyC} (0))
\end{array}
\]









\end{enumerate}

\item If $\G \ot \D \vdash \rlet x = t \rin u\tin \tau$ and $g \in \evalx{{ \G \ot
\D}}$ then we want to show that $q^{\tau}
(\evalx{{\G \ot \D \vdash \rlet x = t \rin u\tin \tau }} \; g) \equiv
close(\rlet x = t \rin u, \G \ot \D, q^{\G \ot \D}(g))\tin \tau$. We will prove
by induction over context definition.

\begin{enumerate}
\item  $\G \ot \D = \emptyC$. Then $g \in \{0\}$.

\[
\begin{array}{l}
q^{\tau} (\evalx{{\emptyC \vdash \rlet x = t \rin u\tin \tau }} \; 0)\\
= q^{\tau} ((g \circ (f \times \mathit{id}) \circ \delta_{\emptyC,\emptyC}) \; 0)\\
= q^{\tau} (g \circ (f(0),\mathit{id}(0)))\\
= q^{\tau} (g (f(0),0))\\
\end{array}
\]

\noindent where

\[
\begin{array}{rcl}
f &=& \evalx{{\emptyC \vdash t \tin \sigma }}\\
g&=& \evalx{{ x \tin \sigma \vdash u \tin \tau}}
\end{array}
\]

Using induction (on the main statement of Lemma 2).

\[
\begin{array}{rcl}
q^{\sigma} (f \; 0)&=& close(t,\emptyC,())\tin \sigma\\
&=& t \tin \sigma
\end{array}
\]

Given $f(0) \in \evalx{{x \tin \sigma}}$,
then the induction gives us:

\[
\begin{array}{rcl}
q^\tau (g \;f(0)) &=& close (u, x \tin \sigma, q^{\sigma}(f(0)))\tin \tau\\
&=& \rlet x = q^{\sigma}(f(0)) \rin u \tin \tau\\
&=&  \rlet x = t \rin u \tin \tau\\
\end{array}
\]

\noindent and

\[
\begin{array}{l}
close(\rlet x = t \rin u,\emptyC,q^{\emptyC}(0))\tin \tau\\
= close(\rlet x = t \rin u,\emptyC, ())\tin \tau\\
= \rlet x = t \rin u \tin \tau
\end{array}
\]










\end{enumerate}

\end{enumerate}

As the work is in progress we show only some cases of the proof.
We are currently working on the entire proof as in the final version of the theory.

\section{Codifying the Category $\Qo$}
\label{app:haskell}

Given a natural number which represents the size of a quantum register
we build the qubit base of right size. For example, the base
for a $2$ qubit quantum register is the list with the four basic values:
[True,True], [True,False], [False,True] and [False,False].

\begingroup\par\noindent\advance\leftskip\mathindent\(
\begin{pboxed}\SaveRestoreHook
\column{B}{@{}l@{}}
\column{10}{@{}l@{}}
\column{E}{@{}l@{}}
\fromto{B}{E}{{}\Varid{base}\in\Conid{Int}\to [\mskip1.5mu [\mskip1.5mu \Conid{Bool}\mskip1.5mu]\mskip1.5mu]{}}
\nextline
\fromto{B}{E}{{}\Varid{base}\;\mathrm{0}\mathrel{=}[\mskip1.5mu [\mskip1.5mu \mskip1.5mu]\mskip1.5mu]{}}
\nextline
\fromto{B}{E}{{}\Varid{base}\;\mathrm{1}\mathrel{=}[\mskip1.5mu [\mskip1.5mu \Conid{True}\mskip1.5mu],[\mskip1.5mu \Conid{False}\mskip1.5mu]\mskip1.5mu]{}}
\nextline
\fromto{B}{E}{{}\Varid{base}\;\Varid{n}\mathrel{=}[\mskip1.5mu \Conid{True}\mathbin{:}\Varid{bs}\mid \Varid{bs}\leftarrow \Varid{bss}\mskip1.5mu]\plus [\mskip1.5mu \Conid{False}\mathbin{:}\Varid{bs}\mid \Varid{bs}\leftarrow \Varid{bss}\mskip1.5mu]{}}
\nextline
\fromto{B}{10}{{}\hsindent{10}{}}
\fromto{10}{E}{{}\mathbf{where}\;\Varid{bss}\mathrel{=}\Varid{base}\;(\Varid{n}\mathbin{-}\mathrm{1}){}}
\ColumnHook
\end{pboxed}
\)\par\noindent\endgroup\resethooks

A vector is implemented as a record which has information about the
size $N$ of the quantum register, and a finite function from the
elements of $base \;N$ to complex probability amplitudes.

\begingroup\par\noindent\advance\leftskip\mathindent\(
\begin{pboxed}\SaveRestoreHook
\column{B}{@{}l@{}}
\column{14}{@{}l@{}}
\column{24}{@{}l@{}}
\column{40}{@{}l@{}}
\column{E}{@{}l@{}}
\fromto{B}{E}{{}\mathbf{type}\;\Conid{Compl}\mathrel{=}{\mathbb C}\;\Conid{Float}{}}
\nextline
\fromto{B}{E}{{}\mathbf{data}\;\Conid{Vec}\mathrel{=}\Conid{Vec}\{\mskip1.5mu size\in\Conid{Int},\Varid{vecfun}\in[\mskip1.5mu \Conid{Bool}\mskip1.5mu]\to \Conid{Compl}\mskip1.5mu\}{}}
\nextline[\blanklineskip]
\fromto{B}{E}{{}\Varid{return}\in[\mskip1.5mu \Conid{Bool}\mskip1.5mu]\to \Conid{Vec}{}}
\nextline
\fromto{B}{14}{{}\Varid{return}\;\Varid{ba}\mathrel{=}{}}
\fromto{14}{E}{{}\mathbf{let}\;\Varid{s}\mathrel{=}\Varid{length}\;\Varid{ba}{}}
\nextline
\fromto{14}{24}{{}\mathbf{in}\;\Conid{Vec}\{\mskip1.5mu {}}
\fromto{24}{E}{{}size\mathrel{=}\Varid{s},{}}
\nextline
\fromto{24}{40}{{}\Varid{vecfun}\mathrel{=}\lambda \Varid{b}\to {}}
\fromto{40}{E}{{}\mathbf{if}\;\Varid{ba}\equiv \Varid{b}{}}
\nextline
\fromto{40}{E}{{}\mathbf{then}\;\mathrm{1}{}}
\nextline
\fromto{40}{E}{{}\mathbf{else}\;\mathrm{0}\mskip1.5mu\}{}}
\ColumnHook
\end{pboxed}
\)\par\noindent\endgroup\resethooks

The function $return$ constructs basic vectors, e.g. $return [False]$
maps $False$ to $1$ and $True$ to $0$.

Linear maps are implemented following the idea that functions mapping
basic values to vectors can be lifted to functions from vectors to
vectors, using the monadic bind combinator.  A linear operator is also
a record storing the size of the input and output quantum registers,
and a function mapping values to vectors (in particular to $vecfun$).

\begingroup\par\noindent\advance\leftskip\mathindent\(
\begin{pboxed}\SaveRestoreHook
\column{B}{@{}l@{}}
\column{17}{@{}l@{}}
\column{E}{@{}l@{}}
\fromto{B}{E}{{}\mathbf{data}\;\Conid{Lin}\mathrel{=}\Conid{Lin}\{\mskip1.5mu \Varid{sizem},\Varid{sizen}\in\Conid{Int},{}}
\nextline
\fromto{B}{17}{{}\hsindent{17}{}}
\fromto{17}{E}{{}\Varid{linfun}\in[\mskip1.5mu \Conid{Bool}\mskip1.5mu]\to [\mskip1.5mu \Conid{Bool}\mskip1.5mu]\to \Conid{Compl}\mskip1.5mu\}{}}
\ColumnHook
\end{pboxed}
\)\par\noindent\endgroup\resethooks

Now we give the behaviour of applying a linear operator to a vector.
To allow overloading for the operator symbol $@\!\!\!>\!\!>=$, we
define some abstract function $@\!\!\!>\!\!>=$ as member of a class
which represents the (monadic) bind, and we say that some domains are an
instance of this class.

\begingroup\par\noindent\advance\leftskip\mathindent\(
\begin{pboxed}\SaveRestoreHook
\column{B}{@{}l@{}}
\column{3}{@{}l@{}}
\column{7}{@{}l@{}}
\column{10}{@{}l@{}}
\column{16}{@{}l@{}}
\column{21}{@{}l@{}}
\column{23}{@{}l@{}}
\column{E}{@{}l@{}}
\fromto{B}{E}{{}\mathbf{class}\;\Conid{Bind}\;\Varid{a}\;\Varid{b}\;\mathbf{where}{}}
\nextline
\fromto{B}{3}{{}\hsindent{3}{}}
\fromto{3}{E}{{}(@\!\!>\!\!>=)\in\Varid{a}\to \Varid{b}\to \Varid{a}{}}
\nextline[\blanklineskip]
\fromto{B}{E}{{}\mathbf{instance}\;\Conid{Bind}\;\Conid{Vec}\;\Conid{Lin}\;\mathbf{where}{}}
\nextline
\fromto{B}{3}{{}\hsindent{3}{}}
\fromto{3}{16}{{}\Varid{v}@\!\!>\!\!>=\Varid{l}\mathrel{=}{}}
\fromto{16}{21}{{}\mathbf{let}\;{}}
\fromto{21}{E}{{}\Varid{s}\mathrel{=}size\;\Varid{v}{}}
\nextline
\fromto{21}{E}{{}\Varid{m}\mathrel{=}\Varid{sizem}\;\Varid{l}{}}
\nextline
\fromto{21}{E}{{}\Varid{n}\mathrel{=}\Varid{sizen}\;\Varid{l}{}}
\nextline
\fromto{21}{E}{{}\Varid{vf}\mathrel{=}\Varid{vecfun}\;\Varid{v}{}}
\nextline
\fromto{21}{E}{{}\Varid{lf}\mathrel{=}\Varid{linfun}\;\Varid{l}{}}
\nextline
\fromto{3}{7}{{}\hsindent{4}{}}
\fromto{7}{E}{{}\mathbf{in}\;\mathbf{if}\;\Varid{s}\equiv \Varid{m}{}}
\nextline
\fromto{7}{10}{{}\hsindent{3}{}}
\fromto{10}{E}{{}\mathbf{then}\;\Conid{Vec}\{\mskip1.5mu size\mathrel{=}\Varid{n},\Varid{vecfun}\mathrel{=}\lambda bs\to {}}
\nextline
\fromto{10}{23}{{}\hsindent{13}{}}
\fromto{23}{E}{{}\Varid{sum}\;[\mskip1.5mu \Varid{vf}\;as*(\Varid{lf}\;as\;bs)\mid as\leftarrow \Varid{base}\;\Varid{s}\mskip1.5mu]\mskip1.5mu\}{}}
\nextline
\fromto{7}{10}{{}\hsindent{3}{}}
\fromto{10}{E}{{}\mathbf{else}\;\Varid{error}\;\text{\tt \char34 Size~mismatch\char34}{}}
\ColumnHook
\end{pboxed}
\)\par\noindent\endgroup\resethooks

An isometry is simply specified as a linear operator and its
application to a vector is given straightforward.

\begingroup\par\noindent\advance\leftskip\mathindent\(
\begin{pboxed}\SaveRestoreHook
\column{B}{@{}l@{}}
\column{5}{@{}l@{}}
\column{16}{@{}l@{}}
\column{E}{@{}l@{}}
\fromto{B}{E}{{}\mathbf{data}\;\Conid{Isom}\mathrel{=}\Conid{Isom}\{\mskip1.5mu \Varid{isom}\in\Conid{Lin}\mskip1.5mu\}{}}
\nextline[\blanklineskip]
\fromto{B}{E}{{}\mathbf{instance}\;\Conid{Bind}\;\Conid{Vec}\;\Conid{Isom}\;\mathbf{where}{}}
\nextline
\fromto{B}{5}{{}\hsindent{5}{}}
\fromto{5}{E}{{}\Varid{v}@\!\!>\!\!>=\Varid{i}\mathrel{=}\mathbf{let}\;\Varid{l}\mathrel{=}\Varid{isom}\;\Varid{i}{}}
\nextline
\fromto{5}{16}{{}\hsindent{11}{}}
\fromto{16}{E}{{}\mathbf{in}\;\Varid{v}@\!\!>\!\!>=\Varid{l}{}}
\ColumnHook
\end{pboxed}
\)\par\noindent\endgroup\resethooks

For instance, some maps from section \ref{sec:catq0} are
defined in our Haskell code as:

\begingroup\par\noindent\advance\leftskip\mathindent\(
\begin{pboxed}\SaveRestoreHook
\column{B}{@{}l@{}}
\column{9}{@{}l@{}}
\column{10}{@{}l@{}}
\column{14}{@{}l@{}}
\column{15}{@{}l@{}}
\column{16}{@{}l@{}}
\column{21}{@{}l@{}}
\column{22}{@{}l@{}}
\column{25}{@{}l@{}}
\column{30}{@{}l@{}}
\column{E}{@{}l@{}}
\fromto{B}{E}{{}\neg \in\Conid{Lin}{}}
\nextline
\fromto{B}{14}{{}\neg \mathrel{=}\Conid{Lin}\{\mskip1.5mu {}}
\fromto{14}{E}{{}\Varid{sizem}\mathrel{=}\mathrm{1},\Varid{sizen}\mathrel{=}\mathrm{1},{}}
\nextline
\fromto{14}{E}{{}\Varid{linfun}\mathrel{=}\Varid{fnot}\mskip1.5mu\}{}}
\nextline
\fromto{14}{16}{{}\hsindent{2}{}}
\fromto{16}{E}{{}\mathbf{where}\;\Varid{fnot}\;[\mskip1.5mu \Varid{x}\mskip1.5mu]\;[\mskip1.5mu \Varid{y}\mskip1.5mu]{}}
\nextline
\fromto{16}{21}{{}\hsindent{5}{}}
\fromto{21}{E}{{}\mid \Varid{x}\not\equiv \Varid{y}\mathrel{=}\mathrm{1}{}}
\nextline
\fromto{16}{21}{{}\hsindent{5}{}}
\fromto{21}{E}{{}\mid \Varid{otherwise}\mathrel{=}\mathrm{0}{}}
\nextline[\blanklineskip]
\fromto{B}{E}{{}\Varid{had}\in\Conid{Lin}{}}
\nextline
\fromto{B}{14}{{}\Varid{had}\mathrel{=}\Conid{Lin}\{\mskip1.5mu {}}
\fromto{14}{E}{{}\Varid{sizem}\mathrel{=}\mathrm{1},\Varid{sizen}\mathrel{=}\mathrm{1},{}}
\nextline
\fromto{14}{E}{{}\Varid{linfun}\mathrel{=}\Varid{had}\mskip1.5mu\}{}}
\nextline
\fromto{14}{15}{{}\hsindent{1}{}}
\fromto{15}{22}{{}\mathbf{where}\;{}}
\fromto{22}{E}{{}\Varid{had}\;[\mskip1.5mu \Conid{True}\mskip1.5mu]\;[\mskip1.5mu \Conid{True}\mskip1.5mu]\mathrel{=}\mathbin{-}\mathrm{1}\mathbin{/}\sqrt{\mathrm{2}}{}}
\nextline
\fromto{22}{E}{{}\Varid{had}\;[\mskip1.5mu \Conid{True}\mskip1.5mu]\;[\mskip1.5mu \Conid{False}\mskip1.5mu]\mathrel{=}\mathrm{1}\mathbin{/}\sqrt{\mathrm{2}}{}}
\nextline
\fromto{22}{E}{{}\Varid{had}\;[\mskip1.5mu \Conid{False}\mskip1.5mu]\;[\mskip1.5mu \Conid{False}\mskip1.5mu]\mathrel{=}\mathrm{1}\mathbin{/}\sqrt{\mathrm{2}}{}}
\nextline
\fromto{22}{E}{{}\Varid{had}\;[\mskip1.5mu \Conid{False}\mskip1.5mu]\;[\mskip1.5mu \Conid{True}\mskip1.5mu]\mathrel{=}\mathrm{1}\mathbin{/}\sqrt{\mathrm{2}}{}}
\nextline[\blanklineskip]
\fromto{B}{E}{{}\mbox{\onelinecomment  receives the total size and size of A}{}}
\nextline
\fromto{B}{E}{{}\Varid{swap}\in\Conid{Int}\to \Conid{Int}\to \Conid{Lin}{}}
\nextline
\fromto{B}{E}{{}\Varid{swap}\;\Varid{n}\;\Varid{m}\mathrel{=}{}}
\nextline
\fromto{B}{9}{{}\Conid{Lin}\{\mskip1.5mu {}}
\fromto{9}{E}{{}\Varid{sizem}\mathrel{=}\Varid{n},\Varid{sizen}\mathrel{=}\Varid{n},{}}
\nextline
\fromto{9}{E}{{}\Varid{linfun}\mathrel{=}{}}
\nextline
\fromto{9}{10}{{}\hsindent{1}{}}
\fromto{10}{25}{{}\lambda \Varid{ba}\to \lambda \Varid{bb}\to {}}
\fromto{25}{30}{{}\mathbf{let}\;{}}
\fromto{30}{E}{{}(\Varid{a},\Varid{b})\mathrel{=}\Varid{splitAt}\;\Varid{m}\;\Varid{ba}{}}
\nextline
\fromto{30}{E}{{}(\Varid{b'},\Varid{a'})\mathrel{=}\Varid{splitAt}\;(\Varid{n}\mathbin{-}\Varid{m})\;\Varid{bb}{}}
\nextline
\fromto{25}{E}{{}\mathbf{in}\;\mathbf{if}\;\Varid{and}\;[\mskip1.5mu \Varid{a}\equiv \Varid{a'},\Varid{b}\equiv \Varid{b'}\mskip1.5mu]{}}
\nextline
\fromto{25}{30}{{}\hsindent{5}{}}
\fromto{30}{E}{{}\mathbf{then}\;\mathrm{1}{}}
\nextline
\fromto{25}{30}{{}\hsindent{5}{}}
\fromto{30}{E}{{}\mathbf{else}\;\mathrm{0}\mskip1.5mu\}{}}
\nextline[\blanklineskip]
\fromto{B}{E}{{}\Varid{diagonal}\in\Conid{Int}\to \Conid{Lin}{}}
\nextline
\fromto{B}{E}{{}\Varid{diagonal}\;\Varid{n}\mathrel{=}{}}
\nextline
\fromto{B}{9}{{}\Conid{Lin}\{\mskip1.5mu {}}
\fromto{9}{E}{{}\Varid{sizem}\mathrel{=}\Varid{n},\Varid{sizen}\mathrel{=}\Varid{n}\mathbin{+}\Varid{n},{}}
\nextline
\fromto{9}{E}{{}\Varid{linfun}\mathrel{=}{}}
\nextline
\fromto{9}{10}{{}\hsindent{1}{}}
\fromto{10}{25}{{}\lambda \Varid{ba}\to \lambda \Varid{bb}\to {}}
\fromto{25}{E}{{}\mathbf{let}\;(\Varid{b1},\Varid{b2})\mathrel{=}\Varid{splitAt}\;\Varid{n}\;\Varid{bb}{}}
\nextline
\fromto{25}{E}{{}\mathbf{in}\;\mathbf{if}\;\Varid{and}\;[\mskip1.5mu \Varid{ba}\equiv \Varid{b1},\Varid{b1}\equiv \Varid{b2}\mskip1.5mu]{}}
\nextline
\fromto{25}{30}{{}\hsindent{5}{}}
\fromto{30}{E}{{}\mathbf{then}\;\mathrm{1}{}}
\nextline
\fromto{25}{30}{{}\hsindent{5}{}}
\fromto{30}{E}{{}\mathbf{else}\;\mathrm{0}\mskip1.5mu\}{}}
\ColumnHook
\end{pboxed}
\)\par\noindent\endgroup\resethooks

}
\bibliographystyle{alpha}
\bibliography{local}

\end{document}
